\documentclass[twocolumn]{aastex631}

\usepackage{color}
\definecolor{red}{rgb}{1,0,0}
\definecolor{green}{rgb}{0,1,0}
\definecolor{blue}{rgb}{0,0,1}

\received{June 13, 2025}
\revised{July 3, 2025}
\accepted{July 3, 2025}

\shortauthors{Correnti et al.}

\begin{document}

\title{FEAST: probing the stellar population of the starburst dwarf galaxy NGC~4449 with JWST/NIRCam}

\author[0000-0001-6464-3257]{Matteo Correnti}
\affiliation{INAF Osservatorio Astronomico di Roma, Via Frascati 33, 00078, Monteporzio Catone, Rome, Italy}
\affiliation{ASI-Space Science Data Center, Via del Politecnico, I-00133, Rome, Italy}

\author[0009-0003-6182-8928]{Giacomo Bortolini}
\affiliation{Department of Astronomy, The Oskar Klein Centre, Stockholm University, AlbaNova, SE-10691 Stockholm, Sweden}

\author[0000-0003-2442-6981]{Flavia Dell'Agli}
\affiliation{INAF Osservatorio Astronomico di Roma, Via Frascati 33, 00078, Monteporzio Catone, Rome, Italy}

\author[0000-0002-8192-8091]{Angela Adamo}
\affiliation{Department of Astronomy, The Oskar Klein Centre, Stockholm University, AlbaNova, SE-10691 Stockholm, Sweden}

\author[0000-0001-6291-6813]{Michele Cignoni}
\affiliation{Dipartimento di Fisica, Università di Pisa, Largo Bruno Pontecorvo 3, 56127, Pisa, Italy}
\affiliation{INAF - Osservatorio di Astrofisica e Scienza dello Spazio di Bologna, via Piero Gobetti 93/3, 40129 Bologna, Italy}
\affiliation{INFN, Largo B. Pontecorvo 3, 56127, Pisa, Italy}

\author[0000-0001-5618-0109]{Elena Sacchi}
\affiliation{Leibniz-Institut für Astrophysik Potsdam (AIP), An der Sternwarte 16, 14482 Potsdam, Germany}

\author[0000-0002-0986-4759]{Monica Tosi}
\affiliation{INAF - Osservatorio di Astrofisica e Scienza dello Spazio di Bologna, via Piero Gobetti 93/3, 40129 Bologna, Italy}

\author[0000-0002-8222-8986]{Alex Pedrini}
\affiliation{Department of Astronomy, The Oskar Klein Centre, Stockholm University, AlbaNova, SE-10691 Stockholm, Sweden}

\author[0000-0002-1832-8229]{Anne S.~M. Buckner}
\affiliation{Cardiff Hub for Astrophysics Research and Technology (CHART), School of Physics \& Astronomy, Cardiff University, The Parade, CF24 3AA Cardiff, UK}

\author[0000-0002-5189-8004]{Daniela Calzetti}
\affiliation{Department of Astronomy, University of Massachusetts, 710 North Pleasant Street, Amherst, MA 01003, USA}

\author[0000-0002-5259-4774]{Ana Duarte-Cabral}
\affiliation{Cardiff Hub for Astrophysics Research and Technology (CHART), School of Physics \& Astronomy, Cardiff University, The Parade, CF24 3AA Cardiff, UK}

\author[0000-0002-1723-6330]{Bruce G. Elmegreen}
\affiliation{Katonah, NY 10536}

\author[0000-0002-2199-0977]{Helena Faustino Vieira}
\affiliation{Department of Astronomy, The Oskar Klein Centre, Stockholm University, AlbaNova, SE-10691 Stockholm, Sweden}

\author[0000-0001-8608-0408]{John S. Gallagher}
\affiliation{Department of Astronomy, University of Wisconsin-Madison, 475 N. Charter Street, Madison, WI 53706, USA}

\author[0000-0002-3247-5321]{Kathryn Grasha}
\altaffiliation{ARC DECRA Fellow}
\affiliation{Research School of Astronomy and Astrophysics, Australian National University, Canberra, ACT 2611, Australia}
\affiliation{ARC center of Excellence for All Sky Astrophysics in 3 Dimensions (ASTRO 3D), Australia}

\author[0000-0003-4910-8939]{Benjamin Gregg}
\affiliation{Department of Astronomy, University of Massachusetts, 710 North Pleasant Street, Amherst, MA 01003, USA}

\author[0000-0001-8348-2671]{Kelsey E. Johnson}
\affiliation{Department of Astronomy, University of Virginia, Charlottesville, VA 22904, USA}

\author[0000-0001-8490-6632]{Thomas S.-Y. Lai}
\affiliation{IPAC, California Institute of Technology, 1200 E. California Blvd., Pasadena, CA 91125}

\author[0009-0009-5509-4706]{Drew Lapeer}
\affiliation{Department of Astronomy, University of Massachusetts, 710 North Pleasant Street, Amherst, MA 01003, USA}

\author[0000-0002-1000-6081]{Sean T. Linden}
\affiliation{Steward Observatory, University of Arizona, 933 N. Cherry Avenue, Tucson, AZ 85719, USA}

\author[0000-0003-1427-2456]{Matteo Messa}
\affiliation{INAF — Osservatorio di Astrofisica e Scienza dello Spazio di Bologna, Via Gobetti 93/3, I-40129 Bologna, Italy}

\author[0000-0002-3005-1349]{Goran Ostlin}
\affiliation{Department of Astronomy, The Oskar Klein Centre, Stockholm University, AlbaNova, SE-10691 Stockholm, Sweden}

\author[0000-0003-2954-7643]{Elena Sabbi}
\affiliation{Space Telescope Science Institute, 3700 San Martin Drive, Baltimore, MD 21218, USA}
\affiliation{Gemini Observatory/NSF's NOIRLab, 950 North Cherry Avenue, Tucson, AZ 85719, USA}

\author[0000-0002-0806-168X]{Linda J. Smith}
\affiliation{Space Telescope Science Institute, 3700 San Martin Drive, Baltimore, MD 21218, USA}

\author[0000-0002-5026-6400]{Paolo Ventura}
\affiliation{INAF Osservatorio Astronomico di Roma, Via Frascati 33, 00078, Monteporzio Catone, Rome, Italy}

\correspondingauthor{M. Correnti}
\email{matteo.correnti@inaf.it}


\begin{abstract}
We present new JWST/NIRCam observations of the starburst irregular galaxy NGC~4449, obtained in Cycle 1 as part of the Feedback in Emerging extrAgalactic Star clusTers (FEAST) program, which we use to investigate its resolved stellar populations and their spatial distributions. NGC~4449 NIR color-magnitude diagrams reveal a broad range of stellar populations, spanning different evolutionary phases, from young main sequence stars, to old red giant branch stars and asymptotic giant branch (AGB) stars. The analysis of their spatial distributions shows that younger ($\leq$ 10 Myr) populations form an S-shaped distribution aligned with the galaxy’s north–south axis, while stars aged 10 -- 60 Myr show shifting concentrations from the north to the south, consistent with the possibility that external interactions or tidal effects may have triggered star formation in spatially distinct bursts. Clusters of comparable ages generally follow these distributions, suggesting that cluster and field stars form at the same pace in each galaxy region. 
Thanks to the unprecedented high-spatial resolution and sensitivity of the JWST data we recover a clear gap between Oxygen-rich and the carbon star branch of the AGB population and the presence of a massive AGB star ``finger''. The analysis of these stars can provide constraints on AGB evolution models and dust production in this galaxy. These results confirms NGC~4449 status as a compelling example of a local dwarf starburst galaxy undergoing complex and possibly external driven star formation and underscore  the power of JWST in probing the full lifecycle of stars in nearby starburst systems.
\end{abstract}



\section{Introduction} 
\label{s:intro}

Star-forming systems are classified as starburst galaxies when they are characterized by extreme 
star formation events, called starbursts, with star formation rates (SFR) from 1 $M_{\odot}/yr$ in relatively normal galaxies to up to 100 -- 1000 $M_{\odot}/yr$ more intense than average in the most extreme cases, over a relatively short duration. Most often starbursts are observed in the central regions of galaxies, but can be found also in peripheral zones, when they are triggered by the effects of interactions with other systems. Starburst galaxies have been found to be frequent at redshift larger than 2 thanks to deep photometric and spectroscopic surveys \citep[e.g.,][]{Steidel1996,Pettini2001,Lefevre2005}, and appear to be the dominant form of star formation in the early universe now revealed by JWST \citep[e.g.,][among many others]{stark2025}.

In the local Universe (within 10 -- 20 Mpc), where individual star and gas features can be resolved, starburst systems are often irregular and blue compact galaxies, either apparently isolated or with observational evidence of companions, which could possibly be the cause of the starburst event \citep[see, e.g.,][for both an observational and a theoretical point of view]{Larson1978,Genzel1998}. 

NGC~4449 is a Magellanic-type irregular galaxy, located at a distance of 3.82 $\pm$ 0.27 Mpc \citep{Annibali2008,Sabbi2018}; it is one of the best-studied and spectacular nearby ($\lesssim$ 5 Mpc)  starbursts, and one of the most luminous and active irregular galaxy in the nearby Universe. NGC~4449 has been observed over a broad range of the electromagnetic spectrum, and shows interesting properties. Its infrared (10 -- 150 $\mu$m) luminosity is $2\times 10^{43}$ erg/s \citep{Thronson1987}, and its integrated magnitude M$_{\rm B}$ = -18.2 makes it $\sim$ 1.4 times as luminous as the Large Magellanic Cloud \citep{Hunter1997}, the prototype of Magellanic irregulars.

\begin{figure*}[thbp!]
\begin{center}
{\includegraphics[width=0.9\textwidth]{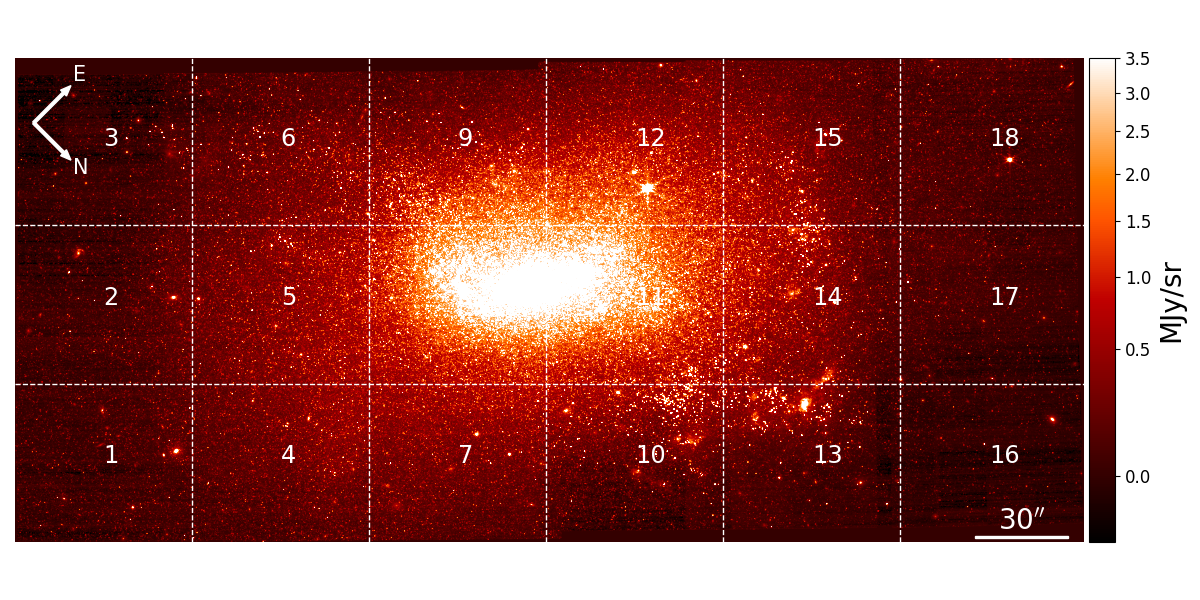}}
\end{center}
\caption{NIRCam F200W NGC~4449 field of view (FoV), corresponding to $\sim$ 2.9 $\times$ 6.35 kpc at the adopted distance. To facilitate the data reduction process we divided the image in 18 tiles (with approximate size of 0.95 $\times$ 1.05 kpc) as indicated by the white dashed lines and numbers in the figure. Angular scale (30$^{\prime\prime}$, corresponding to $\sim$ 550 pc) and North and East directions are also reported.}
\label{fig:fov} 
\end{figure*}

The study of ionized gas in NGC~4449 shows a very turbulent morphology with filaments, shells, and bubbles extending for several kpc \citep{Hunter1990,Hunter1997}. The kinematics of the H~II regions within the galaxy is chaotic, possibly as a consequence of a collision or merger \citep[e.g.][]{ValdezGutierrez2002}. About 40$\%$ of the X-ray emission in NGC~4449 originates from hot gas with a complex morphology, similar to that observed in H$\alpha$, implying an expanding super-bubble with a velocity of about 220 km/s \citep{Summers2003}.
 
At radio wavelengths, NGC~4449 reveals a very extended H~I halo ($\simeq$ 90 kpc in diameter), about ten times larger than the optical diameter of the galaxy, which rotates in the opposite direction of the gas located in the center \citep{Bajaja1994}. \cite{Hunter1998,Hunter1999} have resolved the H~I halo into a central disk-like feature and large gas streamers that surround the galaxy. Both the morphology and the dynamics of the H~I gas suggest that NGC~4449 has undergone relevant interactions in the past \citep{Ai2023}. 

The presence of a gas-rich companion galaxy, DDO~125, at a projected distance of about 40 kpc \citep{Theis2001}, of a diffuse nearby stellar substructure \citep{Martinez-Delgado2012}, and of a massive globular cluster with two intriguing tails of blue stars \citep{Annibali2012} supports the hypothesis that NGC~4449 is affected by external perturbations, e.g. through interactions or mergers with another galaxy, and/or accretion of a gas cloud.

NGC 4449 hosts many star clusters with ages ranging from a few Myr \citep[e.g.,][]{Gelatt2001,Reines2008,McQuaid2024} to older than 9 Gyr \citep[][Pedrini et al., in preparation]{Annibali2018,Whitmore2020}. It features a young ($\sim$ 6–10 Myr) and peculiar central super star cluster (SSC) \citep{Boker2001,Annibali2011} and several other SSCs in the inner region of the galaxy and in other more external star-forming regions \citep{Annibali2011}. Moreover, NGC~4449 exhibits a prominent stellar bar which covers a large fraction of the optical body \citep{Hunter1999}, and a spherical distribution of old (older than 3 -- 5 Gyr) stars \citep{Bothun1986}.

Chemical abundances in NGC~4449 have been derived from its numerous H~II regions \citep[see e.g.,][]{Hunter1982,Talent1989,Pilyugin2015,Annibali2017} and Planetary Nebulae \citep{Annibali2017}, showing an oxygen abundance between $12+log(O/H) = 8.26 \pm 0.09$ and  $12+log(O/H) = 8.37 \pm 0.05$,  and from its many globular clusters \citep{Annibali2018}, which have an iron abundance of $\sim$ -1.2$\leq$[Fe/H]$\leq -0.7$ and sub-solar [$\alpha$/Fe] ratios, with a peak at $\simeq$ -0.4 at epochs older than 9 Gyr ago.

NGC~4449 is close enough that the Hubble Space Telescope (HST) can resolve its individual stars and study in detail its numerous star clusters. This interesting irregular galaxy has been the target of several HST programs, including the HST Treasury program LEGUS \citep{Calzetti2015}, that have led to several studies on its resolved stellar populations \citep[e.g.][]{Annibali2008,Calzetti2015}, star formation history \citep{McQuinn2010,Cignoni2018,Sacchi2018} and star cluster population \citep[e.g.][and references therein]{Annibali2011,Whitmore2020}. 

The current star formation (SF) in NGC~4449 occurs throughout the galaxy \citep{Hunter1997,Cignoni2018,Sacchi2018}, although with different rates from one region to another. With a current SFR of 1 $M_{\odot}/yr$ \citep{Calzetti2015}, NGC~4449 is the star-forming galaxy with the highest recent SFR in the sample
of 18 starburst dwarfs whose star formation history (SFH) was derived by \cite{McQuinn2010}. It is considered similar to Lyman break galaxies at high redshift (z$\simeq$3), where the brightest regions of star formation are embedded in a diffuse nebulosity and dominate the integrated light \citep{Giavalisco2002}. Therefore, studying such nearby galaxies is particularly important, as they can provide valuable insights into the processes of star formation in the early universe.

Leveraging the broad wavelength coverage and superior spatial resolution of JWST, we can conduct, for the first time, a detailed NIR study of the resolved stellar population in NGC~4449. While studies at UV and optical wavelengths are limited by internal dust extinction \citep[see e.g.,][]{Sacchi2018}, JWST observations enable us to pierce through the dust across much of the galaxy's central regions and halo. Thus, our JWST observations covering approximately a region of 2.9 $\times$ 6.35 kpc centered on NGC~4449, facilitate the identification and characterization of stellar populations that are otherwise heavily obscured at shorter wavelengths. These NIR observations are crucial for complementing studies in the UV and optical, enabling a robust comparison across wavelength regimes and providing key constraints on the galaxy's structural and evolutionary properties.

\begin{figure*}[thbp!]
\centering
{\includegraphics[width=0.45\textwidth]{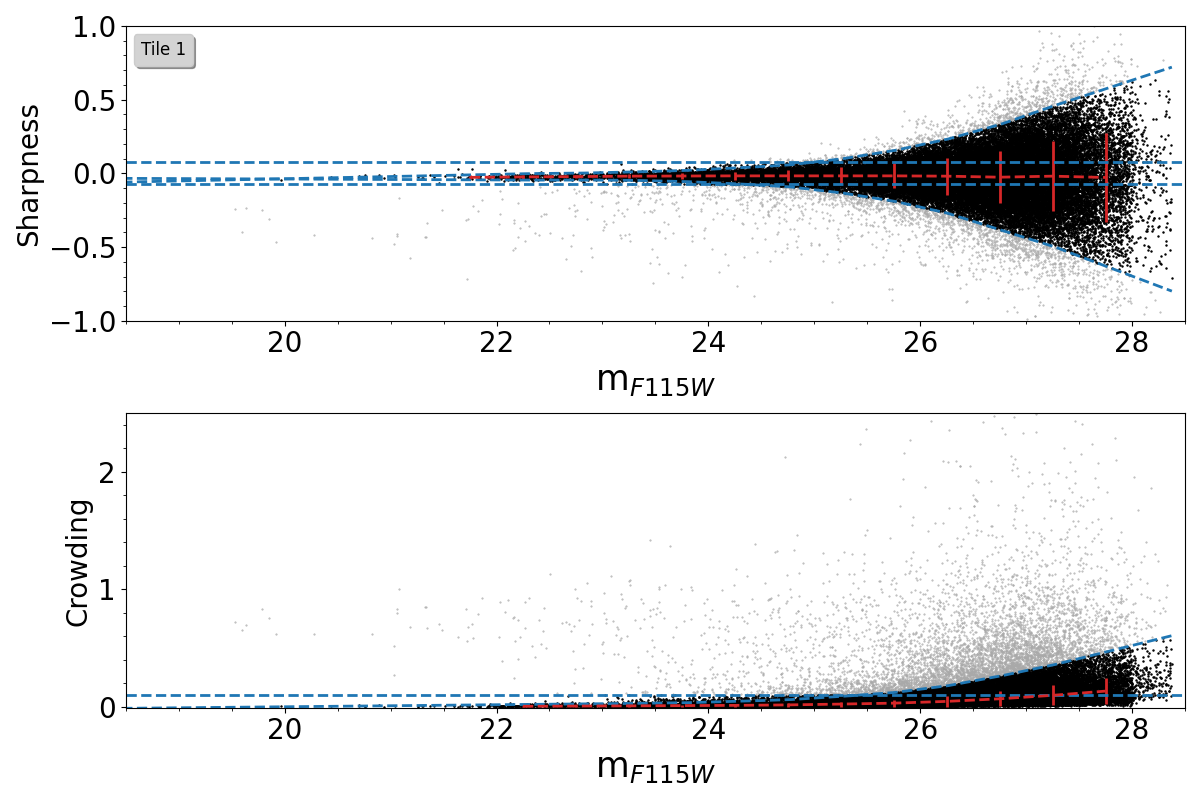}}
{\includegraphics[width=0.45\textwidth]{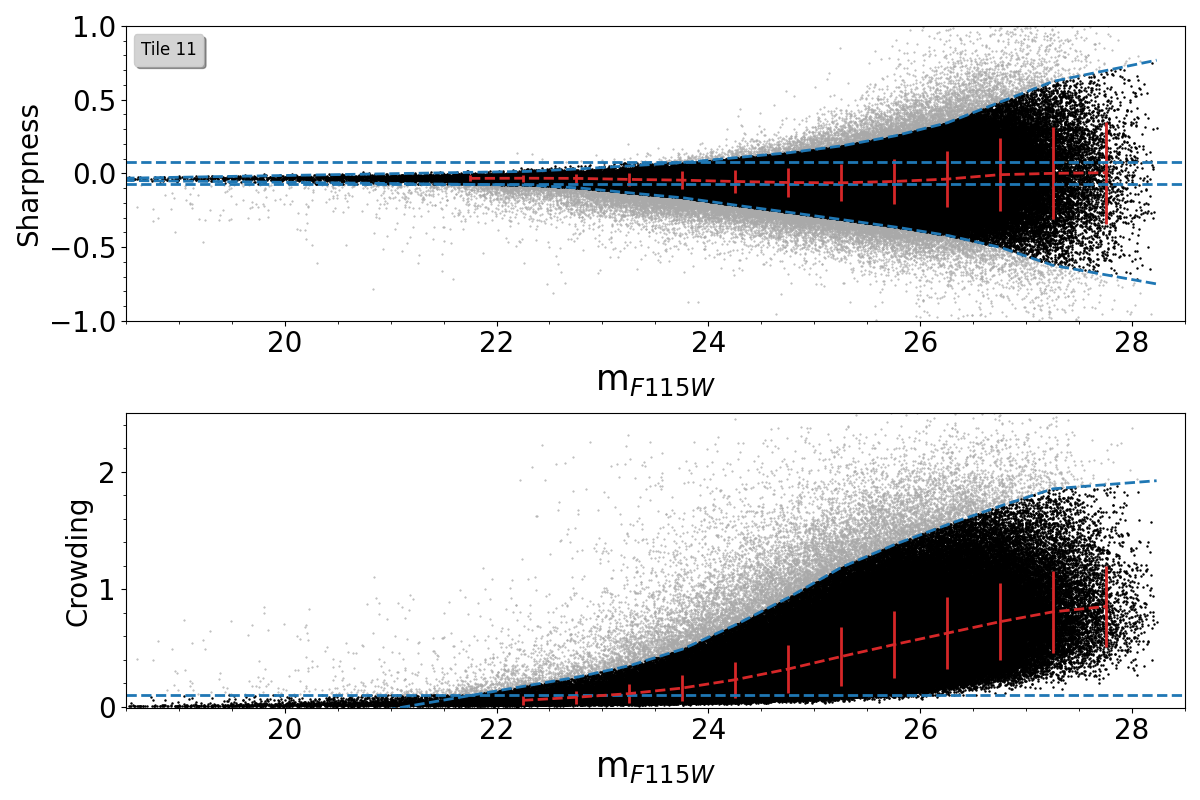}}
\caption{DOLPHOT {\tt sharpness} (top panels) and {\tt crowding} (bottom panels) as a function of m$_{F115W}$ for a tile in the galaxy outskirt (tile 1, left panels) and for a tile covering the central region (tile 11, right panels). Light gray dots represent all the sources passing the preliminary selection cuts (i.e., {\tt Object\_Type}, {\tt quality\_flag}, and {\tt SNR}) while black points indicate the sources that pass {\tt sharpness} and {\tt crowding} cuts. For each bin, we report {\tt sharpness} and {\tt crowding} mean and $\sigma$ and the interpolated values (red dashed lines), whereas blue dashed lines represent the selection limits, as described in Sect.~\ref{s:data_reduction}.}
\label{fig:sel_stars} 
\end{figure*}

In this first paper of a series, we focus on the spatial distribution of NGC~4449's resolved stellar populations, with particular attention to the spatial distributions of young stars formed at different ages and to the analysis of the asymptotic giant branch (AGB) stars morphology and properties. The paper is organized as follows: in Sect.~\ref{s:data_reduction} we describe in detail the data reduction, whereas in Sect.~\ref{s:cmd} we present NGC~4449's color-magnitude diagrams (CMDs) and describe the galaxy stellar content. In Sect.~\ref{s:sp_distr} we describe the spatial distributions of the different age populations and we compare those with the clusters spatial distributions. In Sect.~\ref{s:agb}, we describe the AGB stars color-magnitude and spatial distributions and their properties. Finally, in Sect.~\ref{s:conclusion} we discuss and summarize our results. 

\section{Data Reduction}
\label{s:data_reduction}

NGC~4449 observations have been obtained with the JWST NIRCam and MIRI instruments, as part of the Feedback in Emerging extrAgalactic Star clusTers program (FEAST, GO-1783, PI: Adamo). In particular, NIRCam data have been acquired in filters: F115W, F150W, F187N, F200W, F300M, F335M, F405N, and F444W. For the purpose of this work, we analyzed the data obtained from three SW filters, namely F115W, F150W and F200W, and one LW filter, namely F444W. The two SW filters F115W and F200W represent the best filters combination to study the galaxy resolved stellar population, whereas adding the F150W and F444W filters to the analysis provides the needed wavelength coverage and color combination to study the AGB populations in detail. The total exposure time in the two reference filters is 1116 sec and have been estimated to reach 27 Vegamag in F115W and with a S/N of 3, and 26 Vegamag in F200W with a S/N of 4 enabling us to accurately sample the SFH on average up to 500 Myr in lookback time. These estimates however do not represent the real depth of the data due to crowding effects as discussed in the next section. The total exposure time in the complementary filters is 601 and 1116 sec in F150W and F444W, respectively. Their depth is sufficient to fully sample the NIR bright AGB phases.

Observations were acquired with a FULLBOX 4TIGHT dither pattern. This resulted in deeper coverage in some regions of the mosaic with respect to others. The exposure times were estimated using the shallower regions. The different depth of the data will affect completeness differently and is therefore accounted for in the artificial star test (AST) analysis.

To perform the stellar photometry, we first downloaded from the MAST archive\footnote{\url{https://mast.stsci.edu/portal/Mashup/Clients/Mast/Portal.html}} the calibrated science images ({\it *cal.fits}), which correspond to the output of the second step of the JWST image pipeline, {\it calwebb\_image2}. Similarly to the drizzling process adopted for the HST images, we then run locally the stage three of the JWST pipeline, {\it calwebb\_image3}, in order to create a single-stacked, distortion-corrected image (i.e., {\it *i2d.fits}) that will be used as the reference frame in the photometric data reduction. An intermediate product of the third step of the JWST pipeline are the {\it *crf.fits} images, which are similar to the {\it *cal.fits}, but with absolute astrometric alignment (in our case to GAIA DR3) and cosmic rays flag applied. Those {\it *crf.fits} images were used in the data reduction, while we adopted the F200W stacked image as our reference image. The {\it fits} files have been processed using the pipeline versioning information CAL\_VER = 1.11.3, CRDS\_VER = 11.16.20, and CRDS\_CTX = jwst\_1100.pmap. 

\begin{figure*}[thbp!]
\begin{center}
{\includegraphics[width=0.9\textwidth]{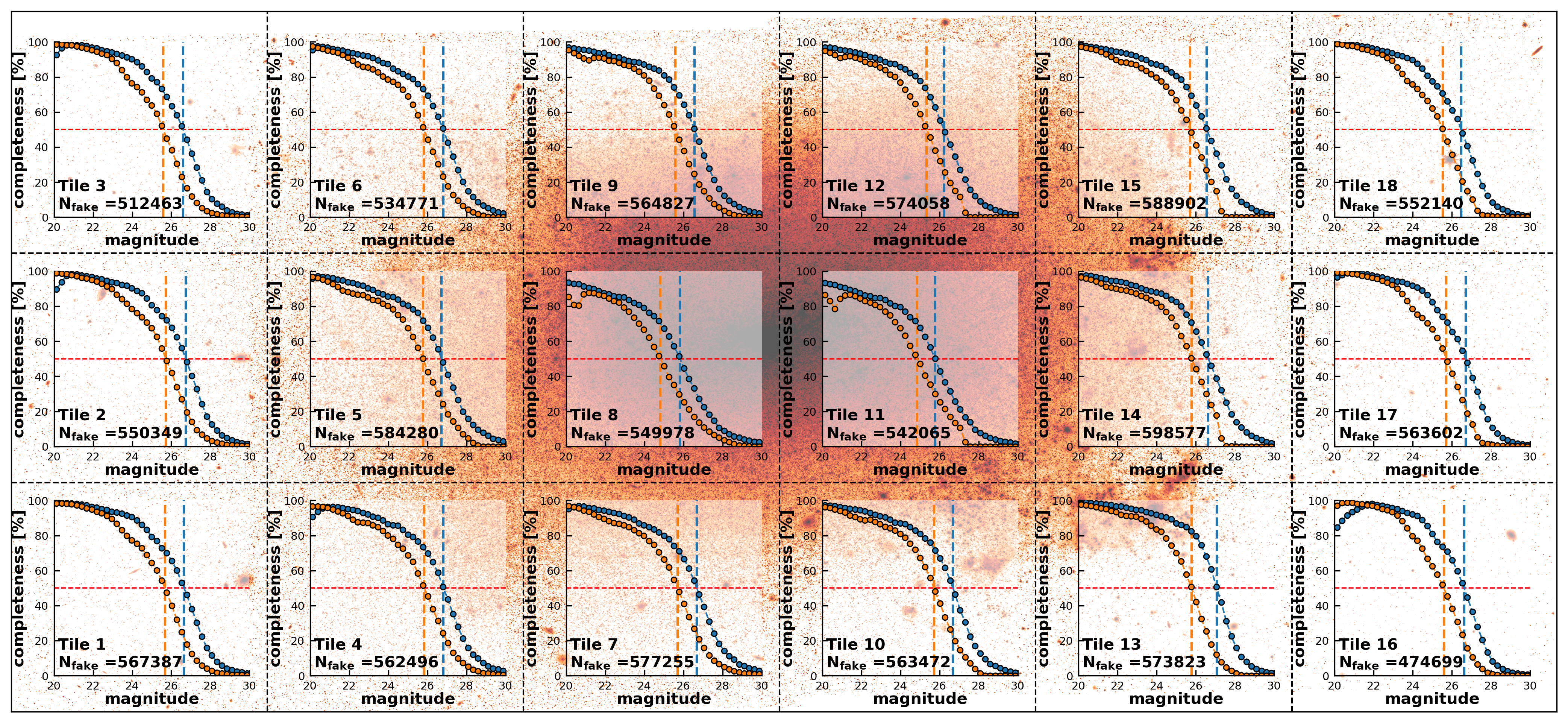}}
\end{center}
\caption{Completeness vs magnitudes (m$_{F115W}$, blue dots; m$_{F200W}$, orange dots) as a function of spatial position in the galaxy (i.e., for each tile), displayed on top of the F200W NGC~4449 image. The number of recovered stars in each tile, out of the 1 million injected, is reported in each panel. 50\% completeness is indicated by the red dashed line, whereas the corresponding m$_{F115W}$ and m$_{F200W}$ are indicated by the blue and orange vertical dashed lines. We note that the drop in completeness at bright magnitudes (i.e., $\sim$ 22 mag) in some tiles, is due to the choice of how the quality cuts are applied to the data, as described in Sect.~\ref{s:data_reduction}.}
\label{fig:completeness} 
\end{figure*}

We used the latest version of the photometry software package DOLPHOT \citep{Dolphin2000, Dolphin2016} with the new beta-version of a specific JWST NIRCam module \citep{Weisz2024} to obtain simultaneous PSF photometry in the F115W, F150W and F200W filters. First, we ran the pre-processing routine {\it nircammask}, specifically designed to perform a series of preliminary steps on the NIRCam images, such as masking out the bad pixels, identifying the saturated pixels, converting the images units from MJy/sr to DN, and applying the pixel area map (PAM) to the single calibrated images. Then, we ran the photometry, setting the DOLPHOT parameters as suggested in \cite{Weisz2024}. Final magnitudes are calibrated in the Vegamag systems, with the updated zeropoints that match the recommended Sirius-Vega based calibration system. 

To simplify and speed up the reduction process, and in particular the AST described below, we decided to divide our reference image into 18 similar-size $\sim$ 0.95 $\times$ 1.05 kpc regions (hereafter tiles) as shown in Fig.~\ref{fig:fov}, and ran the photometry in each one of them independently. 

The final 18 output photometric catalogs were filtered to exclude artifacts and spurious detections using several diagnostic parameters provided by DOLPHOT. In particular, we first selected objects with flag {\tt Object\_Type} $<$ 1, {\tt quality\_flag} $\leq$2, and {\tt SNR} $>$ 2, then we applied further selections using the parameters {\tt sharpness} and {\tt crowding}. The {\tt sharpness} parameter is a measure of how peaked or broad a source is relative to the PSF and helps to reject image artifacts and background galaxies. The {\tt crowding} parameter measures how much brighter a source would have been if stars nearby on the sky had not been fit simultaneously.

\begin{figure*}[thbp!]
\begin{center}
{\includegraphics[width=0.9\textwidth]{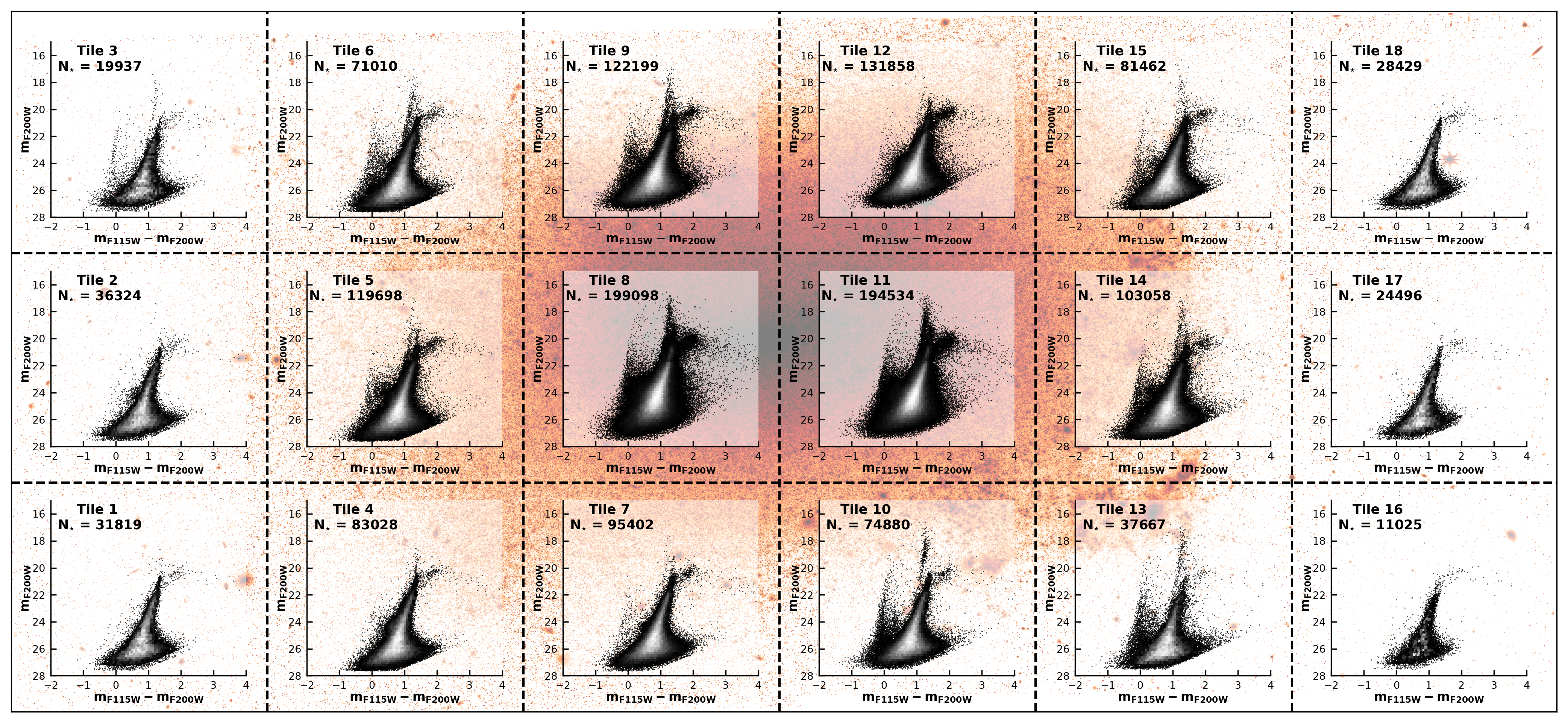}}
\end{center}
\caption{m$_{F200W}$ vs m$_{F115W}$ - m$_{F200W}$ CMDs, plotted as 2D hexagonal bins, as a function of their spatial position, displayed on top of the F200W NGC~4449 image. In each CMD we reported the number of stars. The CMDs of the different tiles clearly show how the presence and density of the different evolutionary phases (e.g., young stars, AGB stars) is strictly correlated to their spatial location within the galaxy.}
\label{fig:cmd} 
\end{figure*}

The {\tt sharpness} cuts have been derived from the {\tt sharpness} vs magnitude distribution, selecting sources with $|${\tt sharpness}$|$ $<$ 0.075 or, after calculating the mean and $\sigma$ of the sharp distribution in 0.5 magnitude  bins, within $\pm 2\sigma$ from the local mean. For what concerns the {\tt crowding} parameter, we selected sources with {\tt crowding} $<$ 0.1 or within $3\sigma$ from the local mean, calculated from the mean and $\sigma$ of the {\tt crowding} distribution in 0.5 magnitude bins. An example of the adopted {\tt sharpness} and {\tt crowding} cuts, for filter F115W, for an outer and an inner tile are reported in Fig.~\ref{fig:sel_stars}. The cuts described above have been applied simultaneously to the m$_{F115W}$ and m$_{F200W}$ photometry. For what concerns the F150W filter, given that the observations are shallower, we did not impose any constraint and we just retained all the sources that pass the other filter cuts, in order to avoid a loss of depth at the CMD faint end. After we applied all the selection cuts, the final catalog (considering all the tiles) contains $\sim$ 1465000 sources (i.e., we retained on average $\sim$ 65\%  of the original photometry sources).  

Given the marginally larger PSF adjustments observed when running simultaneously the SW and LW filters reduction \citep{Weisz2024}, we decided to perform the F444W reduction independently. We adopted the same procedure described above in terms of PSF photometry and catalog culling, and obtained a final catalog containing $\sim$ 270,000 sources. To derive a final combined SW$+$LW catalog, we first selected from the SW catalog only the sources with  m$_{F200W}$ $<$ 23 mag, in order to mitigate the risk of spurious cross-match. This choice is justified by the fact that the target of this combined catalog is the AGB population of NGC~4449, whose stars are much brighter than the imposed magnitude cut. Then, we cross-matched the two catalogs adopting a maximum separation of 0.035 arcsec ($\sim$ half LW pixel), obtaining a final SW$+$LW catalog containing $\sim$ 93.000 sources.

Exploiting our choice of running the photometry in tiles, we run AST in each of them independently. We injected, one at the time, 1 million artificial stars in each tile, producing a total of 18 millions artificial stars. Those stars were distributed in the images following the light distribution to correctly sample the field crowding and with magnitude and color encompassing the observed values in the CMD. Photometry is performed using the same parameters as in the original run and adopting the same selection cuts described above. Stars that are not recovered or do not pass the quality requirements are marked as unrecovered. Fig.~\ref{fig:completeness} shows the m$_{F115W}$ and  m$_{F200W}$ completeness (blue and orange dots, respectively) as a function of magnitude for the 18 tiles, overplotted on the F200W NGC~4449 image. For each tile, we also report the number of fake injected sources that are marked as recovered. As expected, the less crowded outer regions show that the 50\% completeness limit is reached at fainter magnitudes, around $\sim$ 28 mag for m$_{F115W}$ and $\sim$ 26 mag for m$_{F200W}$, whereas the inner region reaches the 50\% completeness limit at m$_{F115W}$ $\sim$ 26 mag and m$_{F200W}$ $\sim$ 25 mag. Conversely, we note that the number of recovered stars is similar across the different tiles. This is due to the varying coverage depths in different parts of the galaxy, caused by the observational setup as described above, combined with the independent selection cuts adopted for each tile, which results in a more conservative selection in the less crowded region, as shown in Fig.~\ref{fig:sel_stars}.

\begin{figure*}[thbp!]
\begin{center}
{\includegraphics[width=0.9\textwidth]{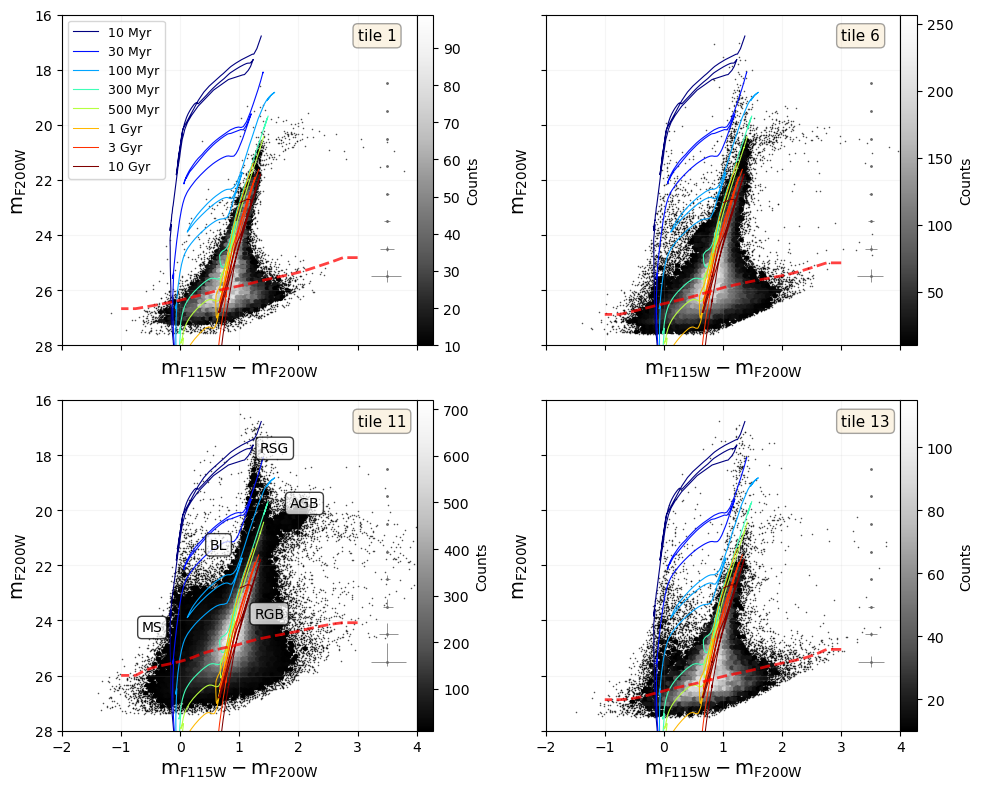}}
\end{center}
\caption{m$_{F200W}$ vs m$_{F115W}$ - m$_{F200W}$ CMDs, plotted as 2D hexagonal bins, for four selected tiles, representative of different regions of the galaxy (Tile 1, outer region, top-left panel; Tile 6 and Tile 13, intermediate regions, right panels; Tile 11, inner region, bottom-left panel). PARSEC stellar isochrones \citep{Bressan2012,Marigo2017,Pastorelli2020} for different ages, reported in the top-left panel, are overplotted with different colors on the CMDs adopting two different metallicities, depending on the age ([Fe/H] = -0.3 dex for isochrones with ages $<$ 1 Gyr, and [Fe/H] = -0.8 dex for ages  $>$ 1 Gyr). Photometric uncertainties and 50\% completeness limit (red dashed line) are also shown. The number of sources for each hexagonal bin are color-coded as indicated by the colorbar adjacent to each panel. The different evolutionary phases, described in Sect.~\ref{s:cmd}, are indicated in the bottom-left panel.}
\label{fig:cmd_select} 
\end{figure*}

\section{Color-Magnitude Diagrams}
\label{s:cmd}

Fig.~\ref{fig:cmd} shows the m$_{F200W}$ vs m$_{F115W}$ - m$_{F200W}$  CMDs of each tile, using the corresponding final catalogs obtained as described above, displayed on top of the F200W NGC~4449 image, similarly to Fig.~\ref{fig:completeness}. A preliminary visual inspection of the CMDs highlights the wide variety of stellar populations present in NGC~4449 and allows us to clearly identify the main features typical of galaxies of this type. Overall, the CMDs display the following features: at m${F115W}$ -- m${F200W}$ $\sim$ 0.0 mag, we observe a blue plume of young (age $<$ 10 – 20 Myr) main sequence (MS) stars, along with core helium-burning stars evolving on the blue side of the characteristic loops traced by evolutionary tracks during this phase, the so-called blue loops (hereafter BL). A red plume appears at m${F115W}$ -- m${F200W}$ $\sim$ 1.1–1.2 mag and m$_{F200W}$ $\leq$ 23.5 mag, corresponding to the cool edge of the BLs. Between the two plumes, stars of intermediate age (up to $\sim$200–300 Myr) trace the full extent of the BL phase. On the bright-end of the CMDs, around m$_{F115W}$ - m$_{F200W}$ $\sim$ 1.5 mag and m$_{F200W}$ $\leq$ 21.5 mag, an almost horizontal tail constituted by thermally pulsing (TP) AGB stars is clearly visible. Finally, a prominent feature is the red giant branch (RGB), at m$_{F115W}$ - m$_{F200W}$ $\sim$ 1.5 mag and m$_{F200W}$ $\geq$ 21.75 mag populated by low-mass stars with ages of at least $\sim$ 2 Gyr. 

However, as is clearly visible in the CMDs of Fig.~\ref{fig:cmd} and already pointed out in \cite{Sacchi2018}, the different populations are not equally distributed in the galaxy. To better appreciate the differences, we selected four tiles, representative of different regions of the galaxy: an outer region (Tile 1), two intermediate regions (Tile 6 and Tile 13, with the latter incorporating part of the bright star forming region present in this part of the galaxy), and an inner region (Tile 11). Their CMDs are shown in Fig.~\ref{fig:cmd_select}.

On top of the CMDs, we display PARSEC-COLIBRI \citep{Bressan2012,Marigo2017,Pastorelli2020} stellar isochrones for ages between 10 Myr and 10 Gyr for two metallicities: [Fe/H] = -0.3 dex for isochrones with ages $<$ 1 Gyr, and [Fe/H] = -0.8 dex for the others. We adopt a distance modulus $(m - M)_0 =$ 28.15 mag \citep{Annibali2008, Tully2013} and reddening $E(B-V) =$ 0.05 mag, derived by optimizing the isochrone fit of the blue edge of the upper MS. In each plot, we also report the photometric uncertainties and the 50\% completeness limit, depicted as a red dashed line. 

\begin{figure*}[thbp!]
\begin{center}
{\includegraphics[width=0.9\textwidth]{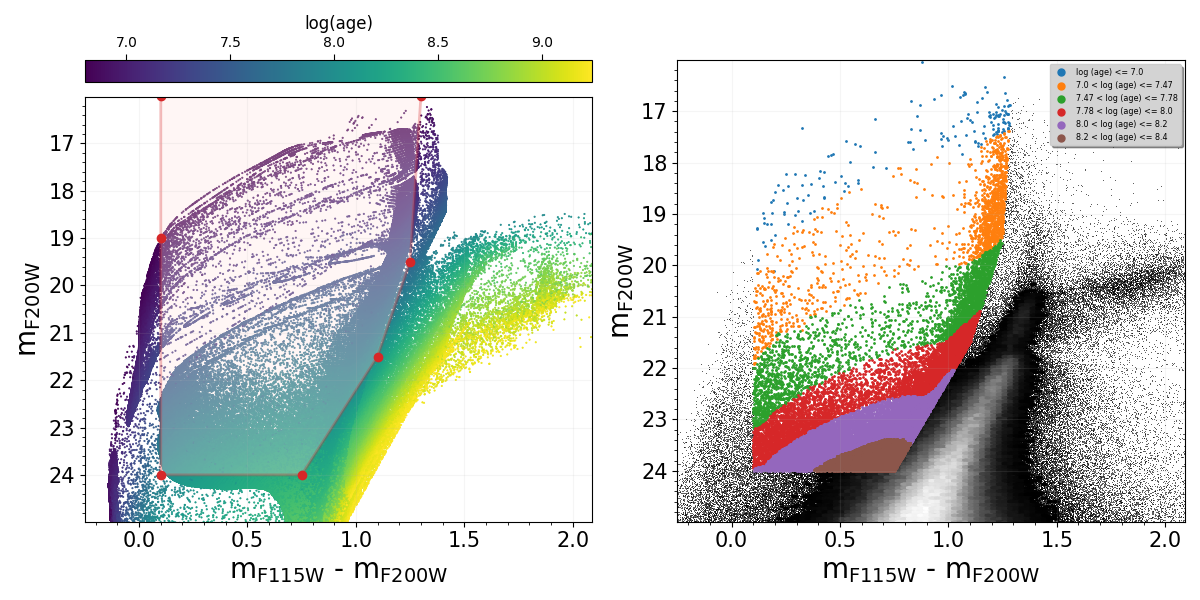}}
\end{center}
\caption{Left panel: Synthetic m$_{F200W}$ vs m$_{F115W}$ - m$_{F200W}$ CMD, obtained as described in Sect.~\ref{s:sp_dist_stars}, with stars color-coded according to their age. The region adopted to analyze the young populations is reported in light-red. Right panel: m$_{F200W}$ vs m$_{F115W}$ - m$_{F200W}$ observed composite CMD with stars color-coded by age intervals (log(age) $\leq$ 7.0, blue dots, 7.0 $<$ log(age) $\leq$ 7.47, orange dots, 7.47 $<$ log(age) $\leq$ 7.78, green dots, 7.78 $<$ log(age) $\leq$ 8.0, red dots, 8.0 $<$ log(age) $\leq$ 8.2, violet dots, 8.2 $<$ log(age) $\leq$ 8.4, brown dots.}
\label{fig:cmd_iso_age} 
\end{figure*}

The CMDs show that, while the RGB is ubiquitously distributed, although with different densities between the inner and outer regions, the same cannot be said for the young stars that populate the blue plume and BLs. In particular, it is evident that in the external regions (e.g., Tile 1), there is a clear lack of very young stars ($<$ 50 Myr), whereas the other tiles show well populated blue plumes. Similar considerations can be done for the BLs at intermediate colors (between the blue plume and red plume): depending on the tile, the ages of the observed population, and the density of stars for each one of them, vary a lot, indicating a clear enhancement of (or lack of) star formation at different epochs for the different regions of the galaxy. A detailed analysis of the SFH of NGC~4449 and the other galaxies observed in FEAST will be presented in a future paper. Finally, we note that the same differences are observed also for the TP-AGB population, although they do not seem to strictly correlate with the location of the tile (see Sect.~\ref{s:agb} for a detailed analysis of NGC~4449 AGB distribution and morphology). 

\section{Spatial distributions}
\label{s:sp_distr}
\subsection{Stellar spatial distributions}
\label{s:sp_dist_stars}

\begin{figure*}[thbp!]
\begin{center}
{\includegraphics[width=1.0\textwidth]{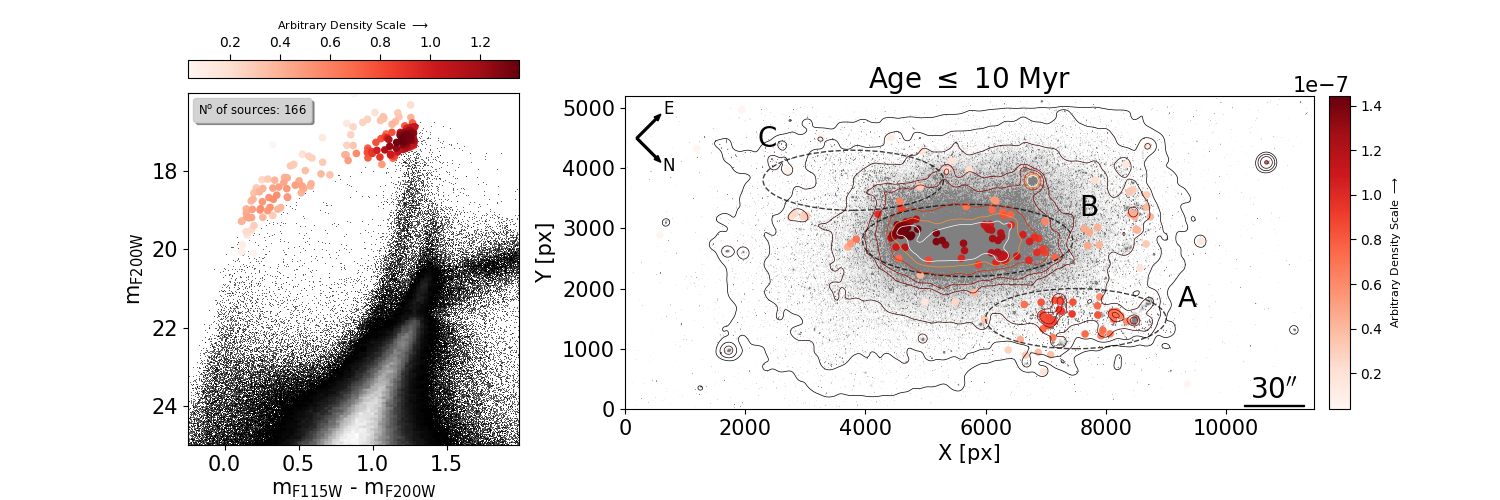}}
{\includegraphics[width=1.0\textwidth]{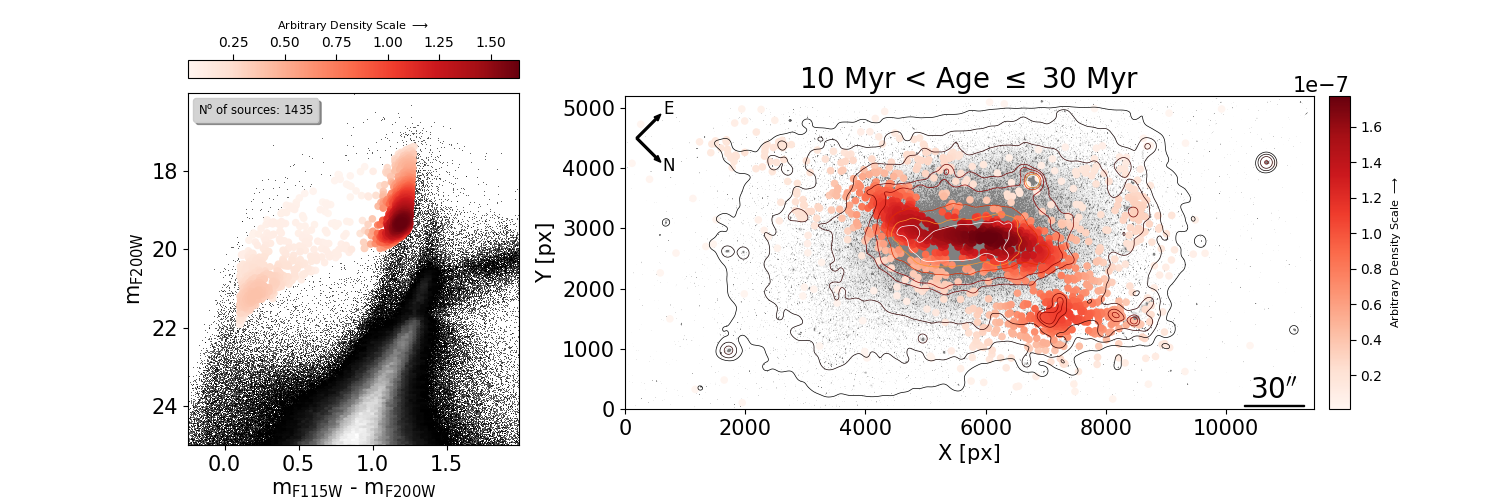}}
{\includegraphics[width=1.0\textwidth]{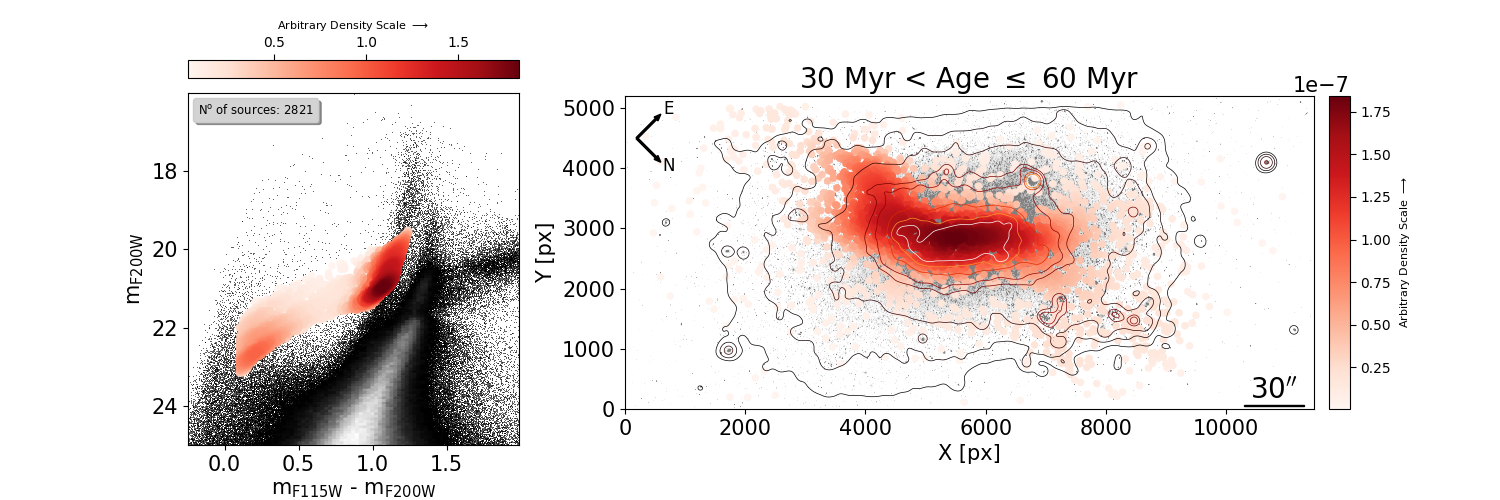}}
\end{center}
\caption{Left panels: m$_{F200W}$ vs m$_{F115W}$ - m$_{F200W}$ composite CMDs, with overplotted in red, color-coded following the star count density, the stars in each specific age intervals (top panel: age $\leq$ 10 Myr; middle panel: 10 Myr $<$ age $\leq$ 30 Myr; bottom panel: 30 Myr $<$ age $\leq$ 60 Myr). In each panel, we report the number of sources. Right panels: corresponding spatial distributions overplotted on the F200W image adopting also in this case a color scale indicating the star count density. In each panel we superimpose to the NGC~4449 image, arbitrary density contours derived from the light distribution in the F200W image. Note that the circular contours observed in different locations (e.g., at X,Y $\sim$ 10500, 4100 px) correspond to bright foreground saturated stars,  star clusters, or galaxies. Angular scale (30$^{\prime\prime}$, corresponding to $\sim$ 550 pc) and North and East directions are also reported. Finally, in the top panel, the three regions discussed in Sect.~\ref{s:sp_dist_stars} are reported as dashed ellipses.} 
\label{fig:sp_distr_young1} 
\end{figure*}

\begin{figure*}[thbp!]
\begin{center}
{\includegraphics[width=1.0\textwidth]{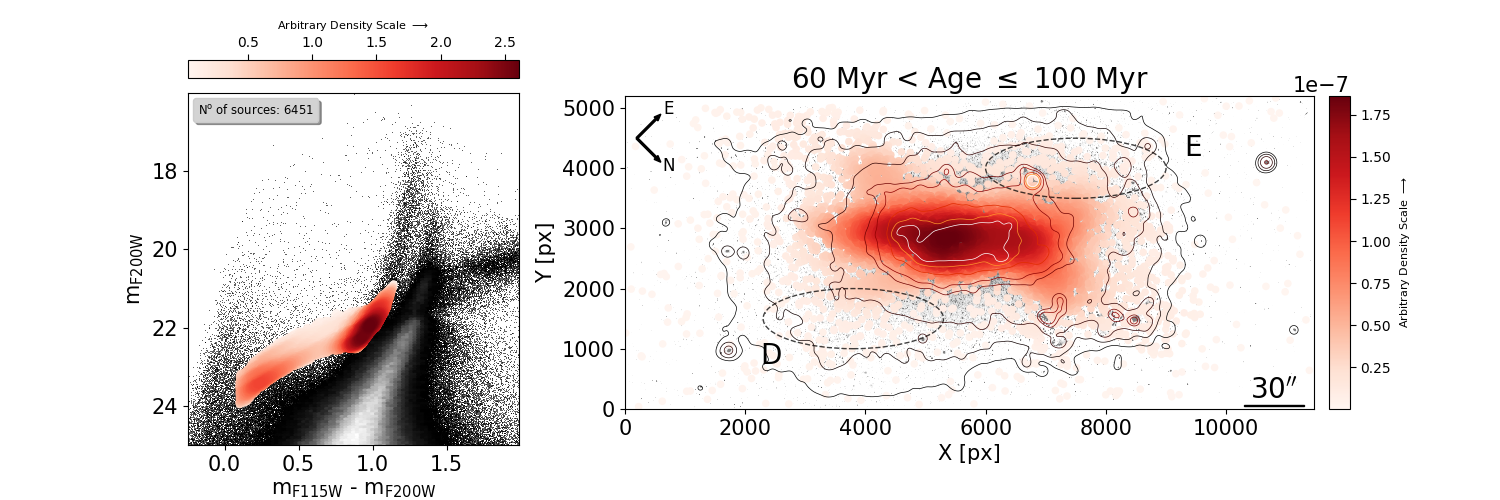}}
{\includegraphics[width=1.0\textwidth]{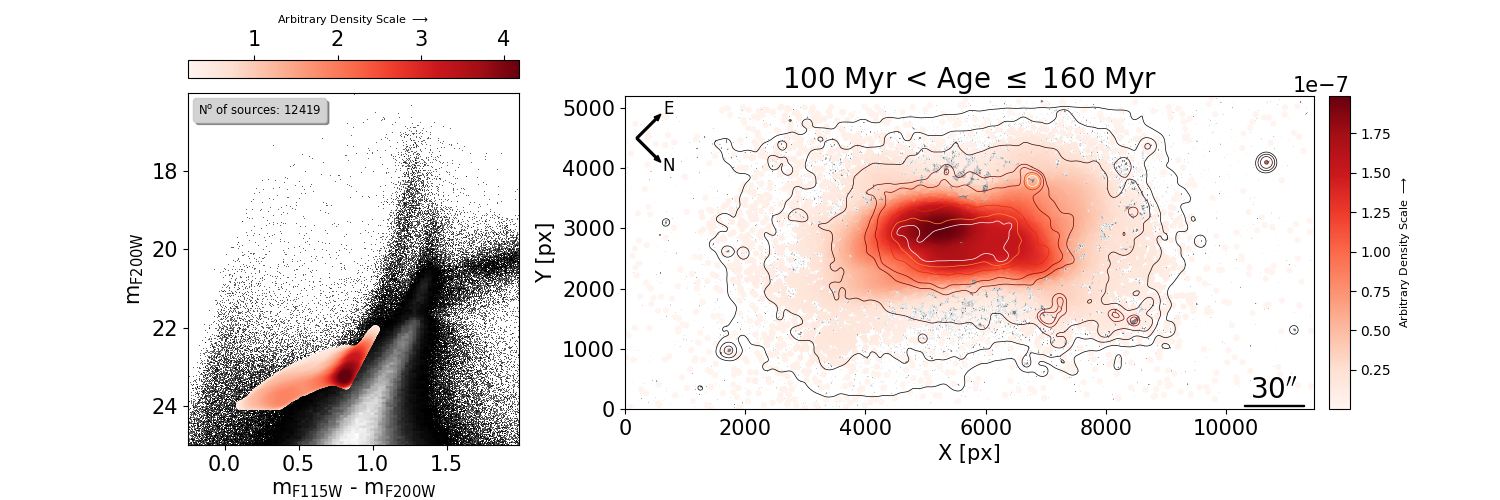}}
{\includegraphics[width=1.0\textwidth]{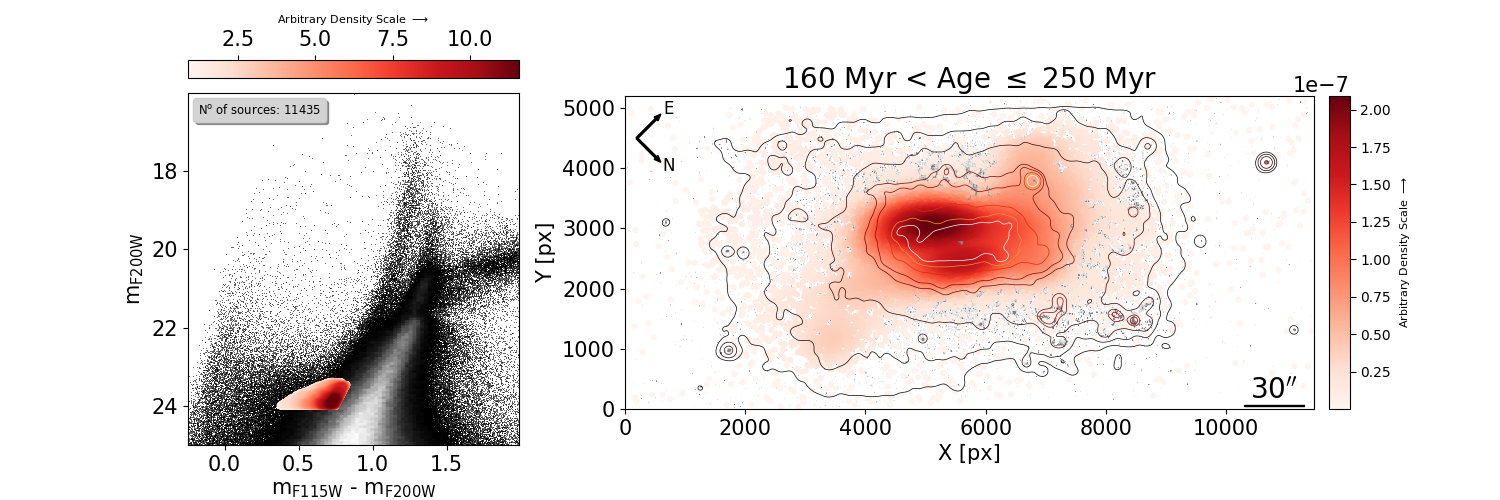}}
\end{center}
\caption{Same as Fig.~\ref{fig:sp_distr_young1} but for the following age intervals: 60 Myr $<$ age $\leq$ 100 Myr (top panel); 100 Myr $<$ age $\leq$ 160 Myr (middle panel); 160 Myr $<$ age $\leq$ 250 Myr (bottom panel). The two regions identifying the western and eastern edges, as described in Sect.~\ref{s:sp_dist_stars}, are reported as dashed ellipses in the top panel.}
\label{fig:sp_distr_young2} 
\end{figure*}

\begin{deluxetable*}{ccccccccc}
\tablecaption{Ratio of star counts in the northern (R1 = A/B), southern (R2 = C/B), western (R3 = D/B), and eastern (R4 = E/B) regions relative to the central region across different age intervals (R3 and R4 derived only for ages $>$ 100 Myr).\label{tab:ratios}}
\tablehead{
\colhead{Age Intervals} & 
\colhead{R1} & \colhead{$\sigma$ R1} & 
\colhead{R2} & \colhead{$\sigma$ R2} & 
\colhead{R3} & \colhead{$\sigma$ R3} & 
\colhead{R4} & \colhead{$\sigma$ R4}
}
\startdata
$\leq$ 10 Myr & 0.453 & 0.093 & 0.004 & 0.023 & -- & -- & -- & -- \\
10 $<$ age $\leq$ 30 Myr & 0.450 & 0.034 & 0.0191 & 0.020 & -- & -- & -- & -- \\
30 $<$ age $\leq$ 60 Myr & 0.129 & 0.011 & 0.287 & 0.017 & -- & -- & -- & -- \\
60 $<$ age $\leq$ 100 Myr & 0.099 & 0.006 & 0.136 & 0.007 & 0.052 & 0.004 & 0.083 & 0.005 \\
100 $<$ age $\leq$ 160 Myr & 0.070 & 0.004 & 0.069 & 0.003 & 0.079 & 0.004 & 0.119 & 0.004 \\
160 $<$ age $\leq$ 250 Myr & 0.032 & 0.002 & 0.062 & 0.003 & 0.127 & 0.005 & 0.169 & 0.006 \\
\enddata
\end{deluxetable*}

To have a clear picture of how NGC~4449 evolved over time, we isolated the different stellar evolutionary phases distinguishable in the CMDs, as described in Sect.~\ref{s:cmd}, and explored their spatial distribution across the galaxy. To achieve this, we followed these steps: first, we created a synthetic CMD. We began by sampling synthetic star masses from a \cite{Kroupa2001} initial mass function (IMF) and assigned ages assuming a constant SFH. Next, we placed the stars on the CMD by interpolating a set of PARSEC-COLIBRI theoretical isochrones \citep{Bressan2012,Marigo2013}. Finally, we transformed the isochrone predictions into magnitude values in the desired filters using the prescriptions provided by \cite{Chen2019}, and assuming distance modulus and reddening appropriate for NGC~4449 \citep{Sacchi2018}. The resulting synthetic CMD is shown in the left panel of Fig.~\ref{fig:cmd_iso_age}, color-coded as a function of the star ages. 
Then, guided by the color-magnitude distribution of stars in the synthetic CMD, we selected an appropriate region in the color-magnitude space to sample the populations we want to analyze. For what concerns the young component, we selected a region in the magnitude range 17 $\lesssim$ m$_{F200W}$ $\lesssim$ 24 mag and in the color range 0.1 $\lesssim$ m$_{F115W}$ - m$_{F200W}$ $\lesssim$ 0.75 mag, at fainter magnitude, that progressively increases towards redder color when moving at brighter magnitudes, in order to fully include the red plume of He-burning stars. The range of the color selection has been chosen such that, on the blue side, the fainter and bluer part of the blue plume is excluded, since it is constituted by stars of different ages, whereas on the red side, we carefully selected our boundary to avoid including in the sample old RGB stars and AGB stars. While this choice provides the best solution to obtain a clear age selection, it is worth to note that, due to the rarity of this phase, it is somewhat affected by stochastic sampling at low SFR (see also Sect.~\ref{s:sp_dist_cl} below). The faint boundary limit at m$_{F200W}$ = 24 mag for the box selecting young core He-burning stars was also set to minimize contamination from different stellar evolutionary phases that at that end tend to merge into the same CMD location at fainter magnitudes. These choices result in an age coverage between a few Myr and $\sim$ 250 Myr. The boundaries of the region selected to study the young populations, and the area encompassed within the points, are indicated by red dots and a light red shade in the left panel of Fig.~\ref{fig:cmd_iso_age}, overplotted on the synthetic CMD.

\begin{figure*}[thbp!]
\begin{center}
{\includegraphics[width=1.0\textwidth]{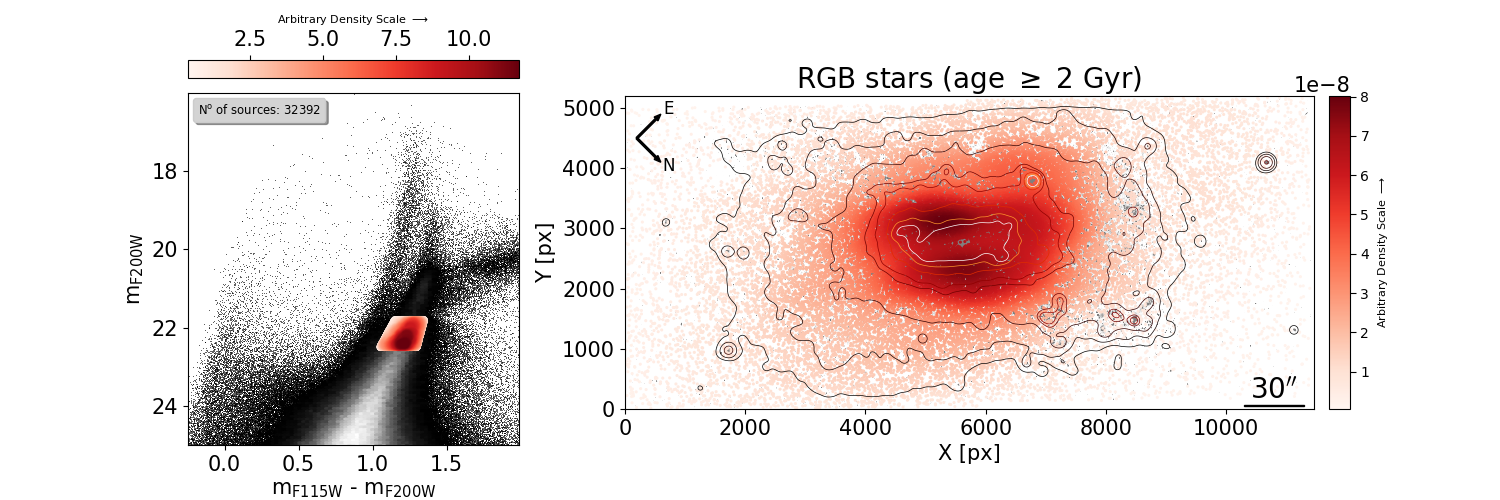}}
\end{center}
\caption{Same as Fig.~\ref{fig:sp_distr_young1} but for RGB stars (age $\geq$ 2 Gyr).}
\label{fig:sp_distr_rgb} 
\end{figure*}

For the observed sources, ages are associated by exploiting the synthetic CMD: for each observed star, we selected all the synthetic stars falling in the same CMD position within a radius of 0.25 mag, and we derived the mean and $\sigma$ of the stars age distribution, which are then assigned to the observed source. If the number of synthetic stars is $<$ 3, we progressively increase the radius size by 0.05 magnitude steps, until we reach the minimum required number of 3 sources inside the radius.

Finally, to derive the spatial distributions within the galaxy of stars in different age bins, we defined a series of age intervals. The choice of the age limits is guided by the need to compromise between having age ranges small enough that the differences between the populations at different ages can be appreciated, but large enough that the results are not hampered by low number statistic and by uncertainties arising from magnitudes photometric errors, at least for populations older than 10 Myr (see discussion below in Sect.~\ref{s:sp_dist_cl}). After different trials, we concluded that a good balance is obtained using the following six age intervals: age $\leq$ 10 Myr, 10 Myr $<$ age $\leq$ 30 Myr, 30 Myr $<$ age $\leq$ 60 Myr, 60 Myr $<$ age $\leq$ 100 Myr, 100 Myr $<$ age $\leq$ 160 Myr, and 160 Myr $<$ age $\leq$ 250 Myr. The sources in each age interval are reported with a different color in the right panel of Fig.~\ref{fig:cmd_iso_age}, which shows the composite CMD for all tiles.

The right panels of Fig.~\ref{fig:sp_distr_young1} and Fig.~\ref{fig:sp_distr_young2} show the different spatial distributions for the six age intervals, overplotted on the F200W image. To guide the reader, in the left panels we display the observed composite CMD where the stars selected in a given age interval are highlighted, color-coded according to the star count density. 

As noted by \citet{Sacchi2018} from the visual HST data, all stars younger than 60 Myr exhibit an S-shaped structure aligned with the bar in the North-South direction, following the H$\alpha$ gas distribution. Interestingly, while both age bins (10 -- 30 Myr and 30 -- 60 Myr) show a similar concentration of stars in the galaxy's center, the recent star formation ($<$ 30 Myr) displays an enhancement in Tile 13 (i.e., the northern side) that is absent in the subsequent age bin. Additionally, the concentration of stars in the 10 -- 30 Myr bin on the opposite side (southern region, around X,Y $\sim$ 4500, 3300 px, Ra,Dec $\sim$ 187.042, 44.083 deg) becomes even more pronounced and extends further outward in the older ($<$ 100 Myr) age bins. This asymmetric configuration is consistent with an external event, possibly a tidal interaction, that triggered star formation in these two regions.

For stars older than 60 Myr, the S-shape progressively fades, with most stars concentrated near the galaxy's center. Notably, for ages above 100 Myr, the peak of the distribution appears slightly offset from the center, as shown in the two bottom panels of Fig.~\ref{fig:sp_distr_young2}. Moreover, in the oldest sampled age bin (160 -- 250 Myr), the S-shape observed at younger ages appears inverted along the East-West direction. 

\begin{figure*}[thbp!]
\begin{center}
{\includegraphics[width=1.0\textwidth]{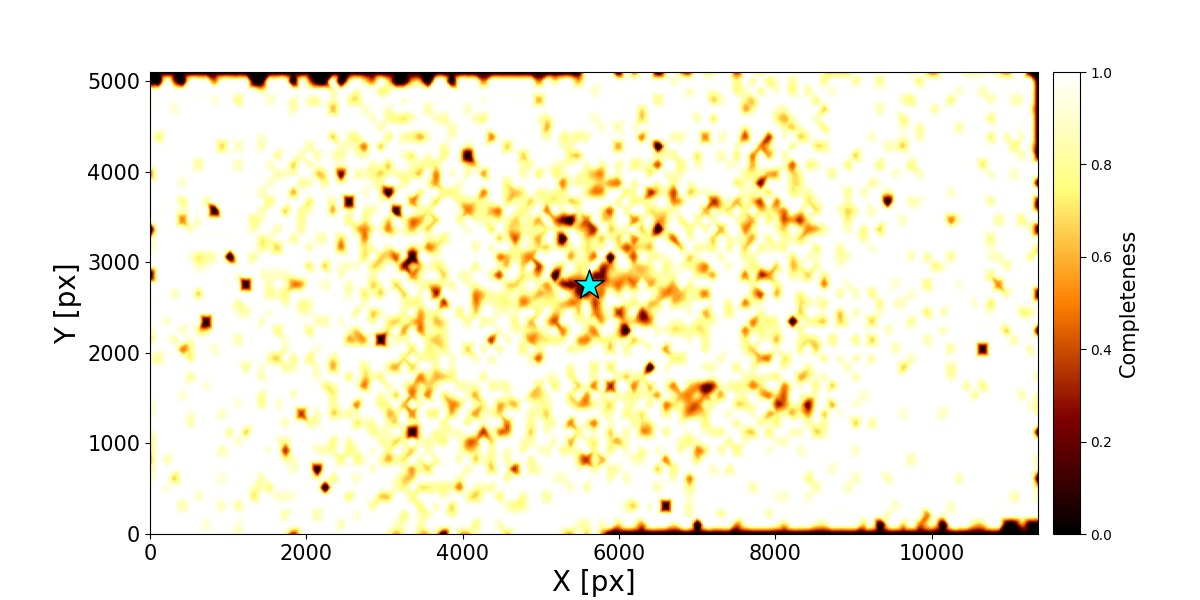}}
\end{center}
\caption{Map of the mean photometric completeness across the FoV, computed in 50$\times$50 px regions within the selected RGB magnitude range. The map reveals a significant drop in completeness coinciding with the central density dip observed in the stellar distribution. The location of the central super star cluster, where saturation effects are more pronounced, is marked with a star.}
\label{fig:sp_comp} 
\end{figure*}

As a preliminary step, we carried out a simple test to better quantify the differences in stellar concentration between the two ends of the North-South S-shaped distribution. A more detailed and rigorous analysis will be presented in the forthcoming SFH paper. In this preliminary approach, we identified three elliptical regions: one centered on the inner central region (referred to as region B) and two covering the northern and southern edges (referred to as region A and region C, respectively). These regions are illustrated with dashed lines in the top panel of Fig.~\ref{fig:sp_distr_young1}. 

We counted the number of stars within each region for different age intervals and computed the ratios of the outer regions to the central one. Specifically, we define R1 = A/B and R2 = C/B, representing the relative stellar densities in the northern and southern parts compared to the center. These ratios, along with their associated uncertainties (estimated from Poisson noise), are reported in Table~\ref{tab:ratios}, and confirm the visual impression. In the youngest age bin ($\leq$ 10 Myr), R1 $\simeq$ 0.45 and R2 $\simeq$ 0.04, indicating a strong concentration of the youngest stars toward the northern edge of the distribution. This northern dominance persists up to $\sim$ 30 Myr, although the southern component begins to increase (R1 $\simeq$ 0.45, R2 $\simeq$ 0.19). Between $\sim$ 30 and 100 Myr, the trend changes: R2 rises significantly and overtakes R1 at $\sim$ 60 Myr (R2 $\simeq$ 0.29, R1 $\simeq$ 0.13), suggesting that the stars formed during this period are more concentrated in the southern part of the system. At older ages (from $\sim$ 100 to 250 Myr), both ratios gradually decline, confirming the progressive fading of the S-shape and a more uniform distribution between the two edges, in particular for the 100 $<$ age $\leq$ 160 Myr intervals.

Similarly, we defined two additional elliptical regions covering the western and eastern edges, referred to as regions D and E, to verify whether the observed inversion of the S-shape is reflected in the star counts. These regions are shown as dashed ellipses in the top panel of Fig.\ref{fig:sp_distr_young2}. Since the inversion becomes evident only for ages $>$ 100 Myr, we derived star counts and ratios only for the three oldest age intervals shown in Fig.\ref{fig:sp_distr_young2}. The resulting ratios, R3 = D/B and R4 = E/B, are reported in Table~\ref{tab:ratios}.

For ages older than 100 Myr, both R3 and R4 show significantly higher values compared to R1 and R2, suggesting that the older stellar populations are more extended along the East–West direction than along the North–South axis. While R1 and R2 drop below 0.1 beyond 100 Myr, R3 and R4 remain higher, reaching up to R4 $\simeq$ 0.17 in the oldest bin.

We also derived the space distribution for the old population (age $\geq$ 2 Gyr) that, as mentioned in Sect.~\ref{s:cmd} from the visual inspection of the different tile CMDs, seems to be ubiquitously distributed in our FoV. To do so, we used the same approach described above, selecting a region in the CMD sampling a portion of the RGB. The choice of the color range and bright magnitude boundary of such region is driven by the isochrones location, whereas for the faint magnitude boundary, we selected stars brighter than m$_{F200W} \sim$ 23.5 mag, in order to have a  completeness above 50\% for all the sources in the different galaxy regions. Selected sources, overplotted on the CMD as a function of star count density, are reported in red in the left panel of Fig.~\ref{fig:sp_distr_rgb}, and their spatial distribution in the right panel. As expected, RGB stars are uniformly distributed, albeit with varying densities between the inner and outer regions of the galaxy. Interestingly, similarly to the young stars aged 160 -- 250 Myr, the RGB stars also exhibit an offset peak relative to the galaxy's center and a double-peak configuration. However, the observed central density dip of RGBs may be attributed to increased crowding and reduced completeness, rather than to an intrinsic feature of the stellar distribution. To test this hypothesis, we divided the FoV into 50$\times$ 50 px regions and computed the mean completeness within our selected RGB magnitude range for each region. The values are shown in Fig.~\ref{fig:sp_comp}, clearly highlighting a drop in completeness that coincides with the observed dip. Furthermore, we note that the dip corresponds to the location of the peculiar central super star cluster, marked as a star in Fig.~\ref{fig:sp_comp}, where the photometry is most severely compromised due to saturation effects.

\subsection{Cluster spatial distributions}
\label{s:sp_dist_cl}

\begin{figure*}[thbp!]
\begin{center}
{\includegraphics[width=1.0\textwidth]{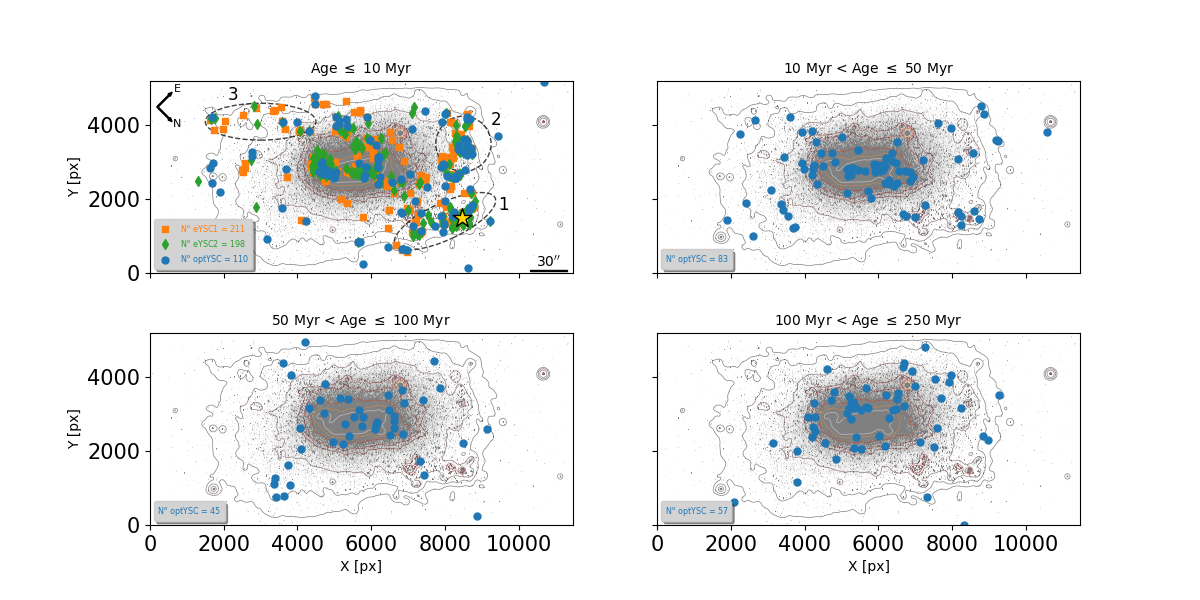}}
\end{center}
\caption{Young star clusters (eYSC1, orange squares; eYSC2, green diamonds, optical YSC, blue dots) spatial distributions as a function of four age intervals (top-left panel: age $\leq$ 10 Myr; top-right panel: 10 Myr $<$ age $\leq$ 50 Myr; bottom-left panel: 50 Myr $<$ age $\leq$ 100 Myr; bottom-right panel: 100 Myr $<$ age $\leq$ 250 Myr). The number of clusters is reported in each panel. For age $\leq$ 10 Myr, the three regions discussed in Sect.~\ref{s:sp_dist_cl} are reported as dashed ellipses and the location of the Wolf-Rayet massive cluster is reported as a gold star.}
\label{fig:sp_distr_clusters} 
\end{figure*}

\begin{figure*}[thbp!]
\begin{center}
{\includegraphics[width=1.0\textwidth]{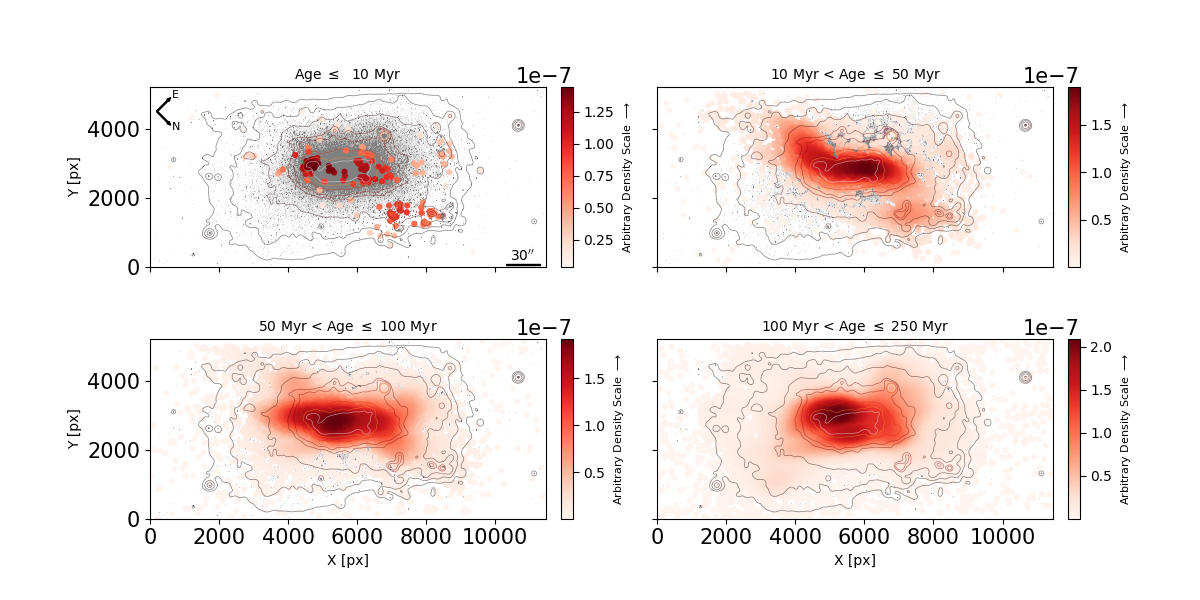}}
\end{center}
\caption{Stellar spatial distributions as a function of the same age intervals adopted in Fig.~\ref{fig:sp_distr_clusters}.}
\label{fig:sp_distr_stars} 
\end{figure*}

We compared the spatial distribution of the resolved young stellar populations in the field with that of young star clusters of the same age. As part of the FEAST project, the identification, selection criteria, classification, and physical properties of the clusters in NGC4449 will be presented in a forthcoming paper (Pedrini et al. subm.). Here, we adopt two different catalogs: for ages $\leq$ 10 Myr, we use the emerging young star cluster (eYSC) catalog \citep[Adamo et al., in preparation, but see also][]{Gregg+24,Pedrini+24} which includes cluster populations largely missed in the UV-optical broadband selection from previous analysis. The emerging young star clusters are divided into eYSCI, containing sources with overlapping peaked emission in both ionized H and  polycyclic aromatic
hydrocarbons (PAHs), bright in star-forming regions, and eYSCII, where the 3.3$\micron$ PAH feature is not peaked anymore, possibly due to the evolution of the surrounding medium due to stellar feedback \citep{Pedrini+24}. We complement this FEAST catalog with the optical young star cluster catalog, based on the LEGUS identified cluster positions and morphological classification to derive an updated photometry (Pedrini et al. subm.). Photometry for the eYSCs and optical catalogs has been estimated using JWST NIRCam data complemented with the HST WFC3 F275 and F336W and  ACS/WFC F435W, F555W, F658N, F814W  observations. The physical properties of the clusters in both catalogs were derived by fitting the observed SED using CIGALE \citep{Boquien+19}. For more details on the fitting process, we refer the readers to  Pedrini et al. subm.

For the spatial distribution of the star clusters, we used broader age intervals with respect to the ones adopted for the stellar populations, due to the low number statistic and a more coarse sensitivity in the cluster age dating method. In particular, while we kept the first age bin as age $\leq$ 10 Myr, for the older bins we increased the age intervals adopting the following values: 10 Myr $<$ age $\leq$ 50 Myr, 50 Myr $<$ age $\leq$ 100 Myr, 100 Myr $<$ age $\leq$ 250 Myr. Fig.~\ref{fig:sp_distr_clusters} shows the clusters spatial distributions (eYSCI, orange squares, eYSC2 green diamonds, optical YSC, blue circles, respectively), whereas Fig.~\ref{fig:sp_distr_stars} shows the stellar spatial distributions for these new age intervals. 

The most meaningful comparison is for the first age bin (age $\leq$ 10 Myr) due to the larger number of clusters. Both distributions exhibit the same enhancement in the northern region, identified as region 1 in the top-left panel of Fig.~\ref{fig:sp_distr_clusters}. This region shows a very active star formation, with the presence of a Wolf-Rayet massive cluster \citep{Sokal+15}, and strong thermal emission, indicating the presence of very young embedded populations \citep{Reines2008}. 

Vice versa, at first glance the clump of youngest clusters located east of this region ((i.e., X,Y $\sim$ 8500, 3500 px --  Ra,Dec $\sim$ 187.076, 44.108 deg, identified as region 2 in Fig.~\ref{fig:sp_distr_clusters})  does not appear to be equally represented in the field star distribution of the top-left panel in Fig.\ref{fig:sp_distr_stars}, although a few young stars are observed there too. Similarly, the young clusters on the southern side of the S-shaped profile (X,Y $\sim$ 3000, 4100 px -- Ra,Dec $\sim$ 187.037, 44.069 deg, identified as region 3 in Fig.~\ref{fig:sp_distr_clusters}) do not seem to have a corresponding counterpart in the stellar distribution of the top-left panel in Fig.\ref{fig:sp_distr_stars}. However, this apparent paucity of field stars with respect to the clusters in the youngest age bin is simply due to our conservative selection of the field stars representative of the age bin. In fact, to avoid as much as possible contamination from stars with different ages falling in the same CMD regions, we restricted our sample to stars in the mid-color portion of the BL phase, excluding stars at its blue-edge and red-edge (see Fig.~\ref{fig:cmd_iso_age}). Taking into account that the central-He-burning phase (although the longest one after the MS phase) lasts only 10$\%$ of the total lifetime of any star, and that most of this time is spent at the blue- and red-edge of the phase, in practice we are sampling only $\sim 1\%$ of the field stars younger than 10 Myr.  When the SFR at the corresponding epoch is not as high as in region 1, we thus deal with small number statistics and can expect 0-1 BL selected stars, when the coeval clusters are of the order of 10-15. In other words, qualitatively the apparent low number of field stars in the age bin is fully consistent with the corresponding number of clusters. A more quantitative comparison will be performed when the SFH is derived in a forthcoming paper.

Regarding the other age bins, the low number statistic across the different age ranges allow us to draw only general conclusions. We observe a good agreement between the two distributions, with the progressive increase of cluster concentration towards the inner region of the galaxy over time. 

\begin{figure*}
\begin{center}
\begin{minipage}{0.45\textwidth}
\resizebox{1\hsize}{!}{\includegraphics{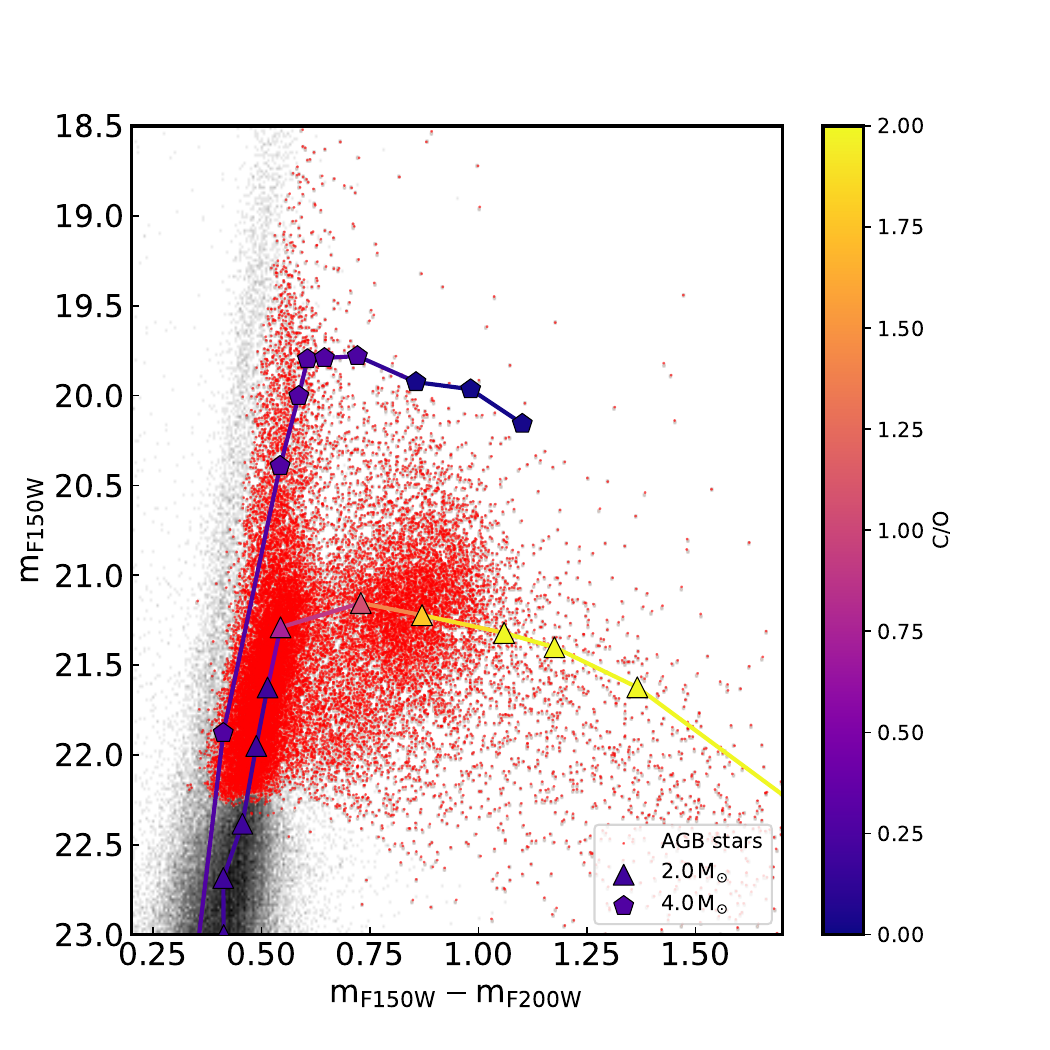}}
\end{minipage}
\begin{minipage}{0.45\textwidth}
\resizebox{1\hsize}{!}{\includegraphics{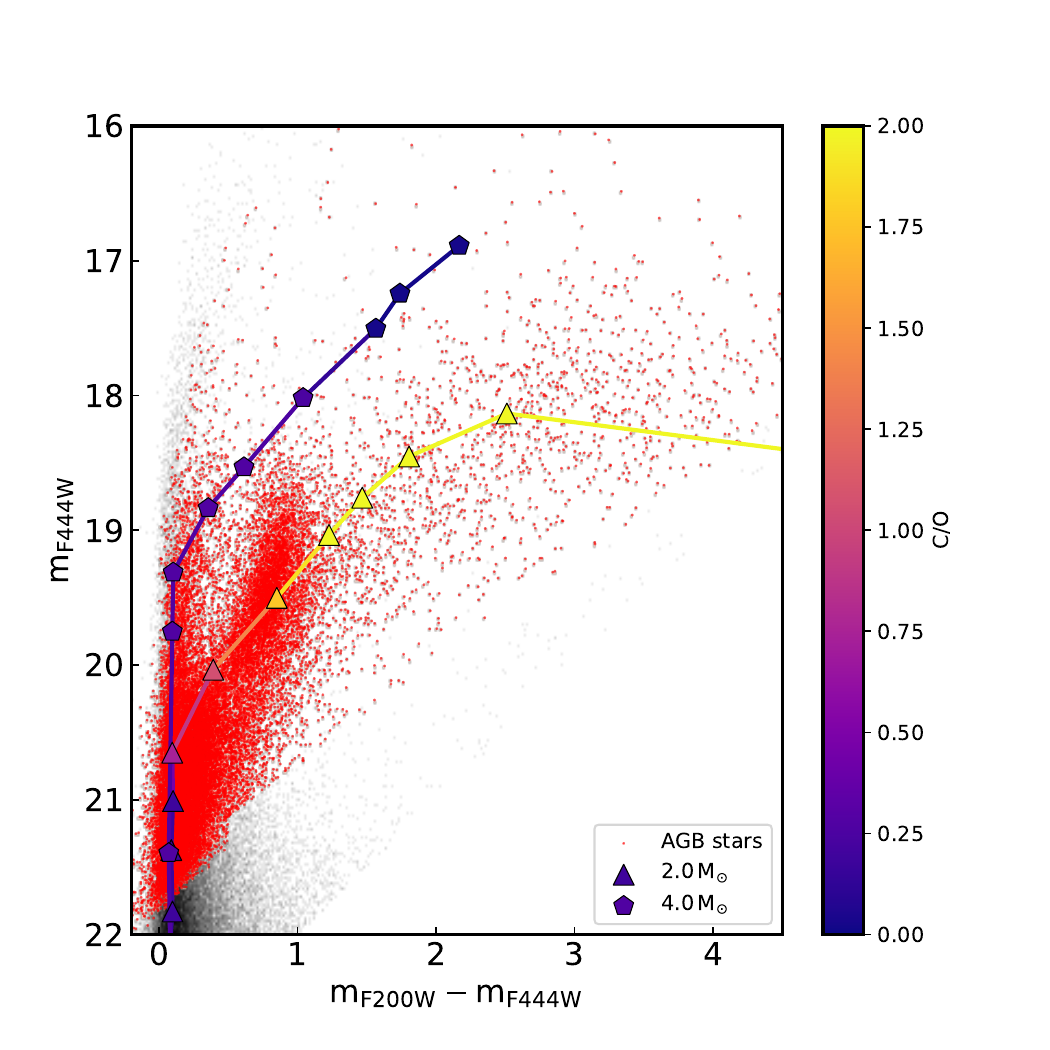}}
\end{minipage}
\caption{NGC 4449 data in the $\rm m_{F150W}\,vs\,m_{F150W} - m_{F200W}$ (left panel) and $\rm m_{F444W}\,vs\,m_{F200W} - m_{F444W}$ (right panel) CMDs, with AGB stars candidates overplotted in red. ATON evolutionary tracks of MS masses 2\,M$_{\odot}$ (triangles) and 4\,M$_{\odot}$ (pentagons) at Z=0.008 are shown in both the panels. The colored bar reports the surface C/O ratio for each model. 
}
\label{fig:cmd_AGB} 
\end{center}
\end{figure*}

\section{AGB stars in NGC 4449}
\label{s:agb}

AGB stars appear in the CMDs as the coolest bright objects, i.e., towards low magnitudes and the reddest colors. They are the progeny of stars with MS masses in the range $\rm 0.8\,M_{\odot} \leq M_{in} \leq 8\,\rm M_{\odot}$, which evolve through the AGB phase after their central helium is exhausted. During this stage, stars expand and lose almost their entire envelope. The chemical composition of the ejecta is altered by internal nucleosynthesis and mixing processes \citep{Karakas14,Ventura22}. High mass-loss rates and low temperatures make the AGB winds ideal for dust formation, with different dust species forming depending on the star's surface composition\citep[e.g.,][]{Ventura14,Nanni13}.  
The oldest AGB stars (with initial masses below $\rm M_{in} \sim 1.2\,M_{\odot}$) evolve as oxygen-rich stars and produce little to no silicate dust. Slightly more massive stars ($\rm 1.2\,M_{\odot} \leq M_{in} \leq 4\,M_{\odot}$), formed between 2 Gyr and 300 Myr ago, undergo enough third dredge-up (TDU) episodes to become carbon stars with C/O $>1$. These stars produce increasing amounts of carbon dust toward the end of the AGB phase. Massive AGB stars ($\rm 4\,M_{\odot} \leq M_{in} \leq 8\,M_{\odot}$) experience hot bottom burning (HBB), where proton capture nucleosynthesis occurs at the bottom of the convective envelope. This process causes them to deviate from the core-mass vs. luminosity relation and prevents them from reaching C/O$>1$ \citep{Paczynski70}. These stars primarily form silicate dust.  
The latter two groups are the dominant dust producers among low- and intermediate-mass stars in the Large Magellanic Cloud \citep{Dellagli15}, a NGC~4449 analogue. The JWST data presented in this work demonstrate an impressive ability to highlight their presence in these near-IR CMDs. Detailed comparisons between the latest generation of AGB+dust models and these data in different CMDs can provide crucial constraints for different assumptions and uncertainties, such as mass loss rate during and before the AGB phase, which determine the evolutionary timescales.

In Fig.~\ref{fig:cmd_AGB}, AGB star candidates are marked in red across the entire NGC~4449 dataset. In the m$_{F150W}$ vs m$_{F150W}$ - m$_{F200W}$ CMD (left panel), AGB stars evolve above the tip of the RGB (m$_{F150W} \sim 22.15$ mag), typically at redder colors than helium-burning stars. To simplify the characterization of the different stars in this region, we overplot the ATON evolutionary tracks \citep{Ventura98}, for which a detailed computation of the AGB phase is coupled with the modeling of dust production in the circumstellar envelope \citep[see, ][for more details]{Ventura12}. In the figure, 2$\rm\,M_{\odot}$ (triangles) and 4$\rm\,M_{\odot}$ (pentagons) models with metallicity Z=0.008, are shown as representatives for different AGB evolutions. The colored shade displays the variation of the surface C/O ratio along the AGB evolution.

The majority of AGB stars are located in the region $\rm 0.3 < m_{F150W} - m_{F200W} < 0.55$ mag, extending in magnitude up to $\rm m_{F150W} \sim 21.5$\,mag, and they can be considered as dust-free, or nearly dust-free. This region is crossed by all AGB stars with MS masses $ > 1.0\,M_{\odot}$, including low-mass stars ($\rm M_{in} < 1.2\,M_{\odot}$) that enter this region towards the end of their AGB evolution. Comparing the number of stars populating this region with the number of stars located within one magnitude below the tip of the RGB can provide a useful constraint to the mass loss experienced during the RGB phase (Dell'Agli in preparation).

At magnitudes around $\rm m_{F150W} \sim 21.2$ mag a bifurcation occurs. As shown by the $\rm 2\,M_{\odot}$ model in the left panel of Fig. \ref{fig:cmd_AGB}, the repeated TDU episodes, which enhance the C/O ratio, lead the stars to enter the carbon star stage (C/O $> 1$). This transition is clearly visible in the CMD in the region $\rm m_{F150W} - m_{F200W} > 0.6$ mag and magnitude $\rm m_{F150W} \sim 21.2$ mag, where the evolution of stars with $\rm 1.2\,M_{\odot} \leq M_{in} \leq 4\,M_{\odot}$ turns towards redder colors and a horizontal carbon star branch appears. These are the AGB stars that produce carbon dust: the more carbon is dredged up, the more copious the dust production becomes, and the redder the colors appear. For this reason, the latest AGB stages of $\rm \sim 2-3\,M_{\odot}$ evolution can be completely enshrouded by dust and invisible at wavelengths $<2-3\, \rm \mu m$. 

We note the presence of a gap between the AGB stars located along the vertical sequence of O-rich stars and those belonging to the carbon star branch. This gap can be explained by an abrupt change in the shape of the spectra, transitioning from an M-type to a carbon star spectrum, with the appearance of different molecular features. Alternatively, it could be due to a rapid transition towards redder colors caused by prolific carbon dust production. A combination of these two effects could also be possible. Either way, this gap should be a common feature in galaxies where a numerous carbon star population is present, although it has not been observed so far in the Magellanic Clouds or other Local Group galaxies, likely because of the poorer resolving power and/or spatial coverage available up to now.

\begin{figure*}[thbp!]
\begin{center}
{\includegraphics[width=1.0\textwidth]{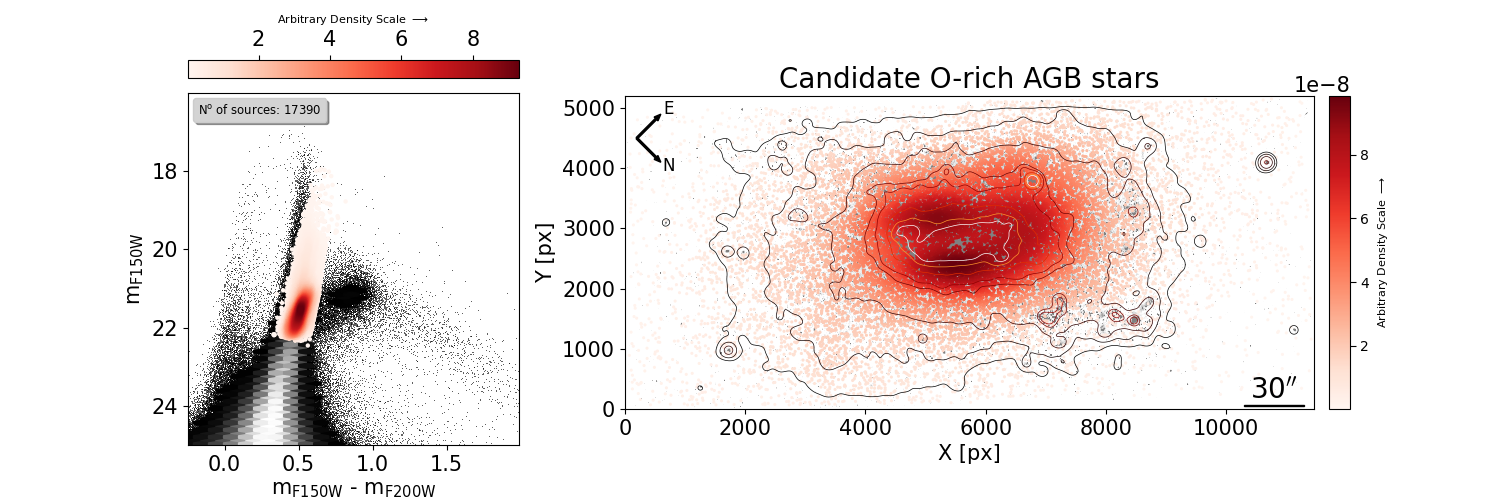}}
{\includegraphics[width=1.0\textwidth]{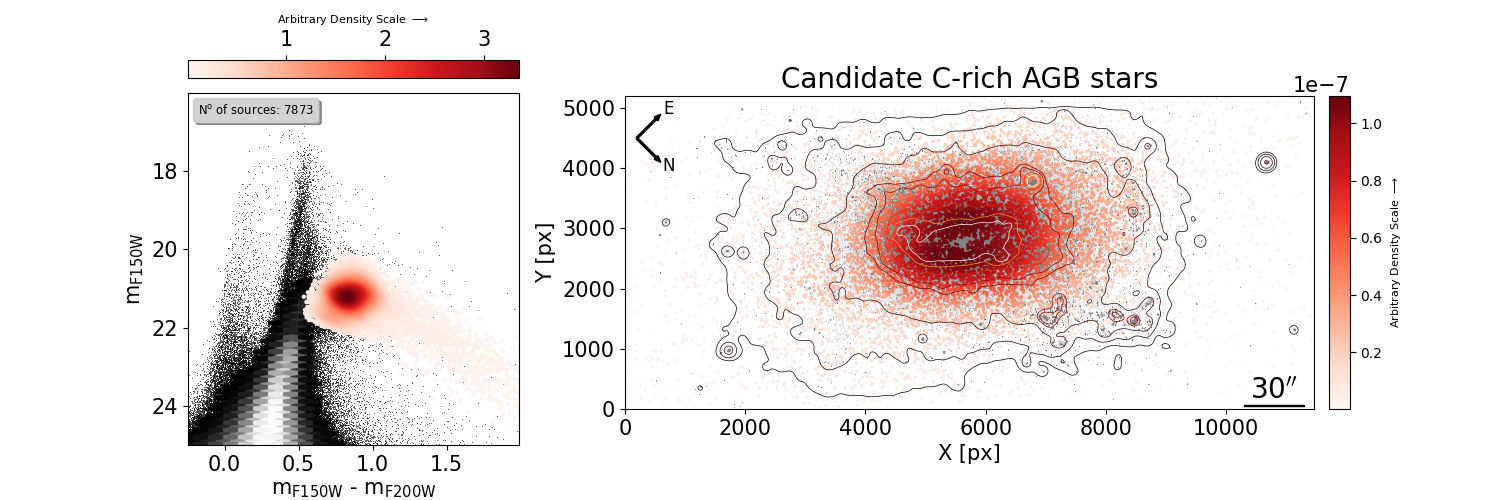}}
\end{center}
\caption{Same as Fig.~\ref{fig:sp_distr_young1} but for the candidate O-rich AGB stars (top panels) and candidate C-rich AGB stars (bottom panels).}
\label{fig:sp_distr_agb} 
\end{figure*}

In Fig.~\ref{fig:cmd_AGB}, we note a significant presence of AGB stars that are brighter than the carbon stars. With regard to these stars, it is interesting to note two drops in the luminosity function. The first, at $\rm m_{F150W} \sim 21$ mag, corresponds to the region where massive AGB stars separate from the carbon stars. The $\rm 4\,M_{\odot}$ model, representative of this class of stars, evolves vertically along the M-type star sequence up to $\rm m_{F150W} \sim 19.7$ mag, where the second drop occurs. At this magnitude, the HBB is activated, favoring brighter configurations and higher mass-loss rates, which in turn promote dust production \citep{Dellagli15,Ventura12}. This is the reason why the star moves towards redder colors. The height of the massive AGB star "finger" depends on several factors, including the IMF and the specific star formation history of the galaxy, which could influence the formation of stars more massive than $\rm 4\,M_{\odot}$. However, where a significant SFR occurs between 30 and 200 Myr ago, as in the case of NGC~4449, the massive AGB stars' finger could serve as a clear constraint for the mass-loss prescription in massive AGB stars. In fact, we expect that the magnitude at which massive AGB stars turn redder is tightly correlated with the adopted mass-loss rate \citep[see e.g.][]{Marini23}. In the models shown in Fig.~\ref{fig:cmd_AGB}, \citet{Bloecker95} mass loss rates are adopted for massive AGB stars' evolution during the O-rich phases \citep{Ventura12}. The strong dependence of Bl\"ocker's mass-loss rate on the star's luminosity favors significant dust production at early stages of the AGB evolution. Under these conditions, massive AGB stars turn towards redder colors as soon as they reach luminosity higher than 20000$\,\rm L_{\odot}$. When different mass loss prescriptions are adopted, such as those of \citet{vw93}, the peak of the mass loss rate and of dust condensation is reached towards the end of the AGB phase, when the stars assume the most expanded configuration, and the pulsation period is at its maximum \citep[see e.g.][for more details]{Marini23,Ventura18}. Therefore, in the latter case we expect a longer evolution and as a consequence a brighter evolution in the vertical sequence with respect to the Bl\"ocker's mass loss case.
To understand what is behind both the "carbon branch gap" and the height of the massive AGB stars' finger, a more quantitative analysis is needed, which is beyond the scope of this paper but is currently ongoing (Del'Agli et al., in prep.).

In the right panel of Fig.~\ref{fig:cmd_AGB}, AGB star candidates are shown in the m$_{F444W}$ vs m$_{F200W}$ - m$_{F444W}$ CMD. This diagram clearly displays the full extent of the carbon star branch, reaching the reddest AGB stars. The spectra of the dustiest stars peak at wavelengths greater than 2$\,\rm \mu m$, while they appear fainter in the other CMD presented here. Conversely, the vertical sequence of O-rich AGB stars overlaps with helium-burning stars in this diagram, making a clean selection of AGB candidates difficult. This highlights the importance of using both color combinations to study AGB stars effectively. 

Finally, in Fig.~\ref{fig:sp_distr_agb}, we show the spatial distributions of the candidate O-rich and C-rich AGB stars. These candidates were manually selected using the CMD combinations presented in Fig.~\ref{fig:cmd_AGB}. We adopted a conservative approach to obtain the cleanest sample possible, excluding sources on the He-burning branch and stars below the tip of the RGB. While this method minimizes contamination from other evolutionary phases, it inherently results in an incomplete sample; therefore, we refrain from drawing strong conclusions based on the spatial distributions. In general, the candidate O-rich AGB stars display a uniform distribution, with density varying as a function of distance from the galaxy center —- similar to the distribution observed for the RGB stars. In contrast, the candidate C-rich AGB stars appear slightly more concentrated toward the inner region.

\section{Discussion and Conclusions}
\label{s:conclusion}

In this work, we presented new JWST/NIRCam observations of the nearby starburst galaxy NGC~4449, obtained as part of the FEAST program. Our primary aim was to examine the galaxy's resolved stellar populations and to explore their spatial distributions. By doing so, we gained insights into the processes governing NGC~4449's star formation episodes and the interplay between its field-star population and its rich young star cluster system. This study serves as the first step towards a more comprehensive analysis that will include deriving the galaxy's SFH in detail and comparing it to the SFH obtained from visible-light data. 

The derived CMDs confirm that NGC~4449 hosts a variety of stellar populations formed over a broad timescale. We detect very young stars ($\leq$ 10 Myr) forming a pronounced blue plume, older helium-burning stars (tens of Myr to few hundred Myr) occupying intermediate colors, and a well defined RGB populated by old low-mass stars at least older than a couple of Gyr. The galaxy also shows a conspicuous presence of AGB stars, with a clear separation between O-rich and C-rich stars, confirming the fundamental impact of high-resolution IR observations to perform a detailed study of such objects. 

The spatial distribution of the youngest stellar populations ($<$ 60 Myr) follows an S-shaped morphology extending along the north–south axis. This structure is consistent with the distribution of ionized gas, suggesting that recent star formation may have been fueled or shaped by gas flows and feedback processes. The S-shape is especially prominent in the 10 -- 30 Myr age bins, after which the northern enhancement fades and new concentrations of young stars appear toward the south. Such a pattern may reflect sequential star formation events triggered by internal feedback or external interactions. In fact, while local conditions such as gas density, turbulence, or mechanical feedback from stellar winds and supernovae play an important role in the spatial distribution of star formation, the effect of interaction and gas accretion might be dominant for NGC~4449 in triggering bursts at different locations. In this context, it is interesting to note that H~I observations of the extended gas component presented in \citet{Hunter1998} reveal a central condensation of gas within a region of approximately 9 kpc, embedded in an elongated elliptical structure. This structure shows a higher concentration toward the northeastern end and features a long gaseous streamer extending from the eastern side. Using a set of N-body simulations, \citet{Theis2001} were able to reproduce the observed H~I morphology of NGC~4449 by assuming a past close encounter with DDO~125, with the closest approach occuring around $\sim$ 400 -- 600 Myr ago. Given that our FoV lies entirely within the inner ellipsoidal region, it is difficult to directly assess whether the observed S-shaped profile is connected to the extended stream-like distribution of the gas. However, it is plausible that both features are effects of the same interaction-driven processes. 

Furthermore, \citet{Lelli2014} demonstrated that enhanced activity of star formation in starburst galaxies is often associated with disturbed H~I morphology, suggesting that such bursts are more likely triggered by external mechanisms than by internal dynamics. In addition to the interaction with DDO~125, the discovery of a stellar tidal stream from a disrupted dwarf galaxy in the halo of NGC~4449 \citep{Martinez-Delgado2012} and the identification of a possible gas-rich accreted satellite remnant  \citep{Annibali2012} further support the scenario in which external accretion events have played a major role in shaping the recent dynamical and star formation history of the galaxy. 

For stellar populations older than 100 Myr, the S-shaped structure becomes less pronounced, and the stars appear more centrally concentrated. There is also a possible indication of an inversion in the orientation of the S-shape profile, from a North–South alignment to an East–West one, for stars older than approximately 150 Myr. Such shift in orientation could potentially be associated to shear effects. However, our field of view encompasses only the inner region of the galaxy ($\sim$ 5$^{\prime}$), which is characterized by a rigid (i.e., solid-body) rotation curve \citep{Bajaja1994}, where the rotational velocity increases linearly with radius. In such a regime, the influence of shear is expected to be minimal due to the lack of significant differential rotation, making it unlikely that shear is responsible for the observed inversion. Conversely, although the observed variation concerns the projected distribution rather than a direct measurement of stellar rotation, the shift in orientation is nonetheless suggestive of an evolving dynamical structure, potentially shaped by tidal interactions. This scenario is particularly interesting in light of previous observations indicating the presence of counter-rotating gas in NGC 4449 (Bajaja et al. 1994; Hunter et al. 1998). These studies show that the ionized and neutral gas rotate in the opposite direction to the main stellar body, with a kinematic major axis oriented approximately East–West, contrasting with the North–South orientation observed in the bulk of the older stellar component. The spatial alignment of the younger ($\leq$ 50 Myr) stars along the North–South axis, broadly consistent with the orientation of the oldest stellar component, may reflect recent star formation occurring in a region that has dynamically settled. In contrast, the slightly older population ($\geq$ 150 Myr), which still shows an East–West alignment, may preserve the spatial imprint of the counter-rotating gaseous structure from which it formed. While the observed distributions do not directly trace stellar rotation, the evolving orientation of the stellar populations supports a scenario in which recent accretion or merger events have introduced kinematically misaligned gas, subsequently forming stars with distinct spatial, and potentially dynamical, properties. This complex configuration highlights the presence of multiple, possibly decoupled, dynamical components in the central regions of NGC 4449. 

Finally, we also note possible offset-center distributions in the oldest ($>$ 160 Myr) bin and in the RGB/AGB sample, although this feature might be due to crowding effects in the latter case. The densest central regions of NGC~4449 are challenging for photometry, and crowding can create apparent `gaps' in the star counts. At any rate, these older stars are clearly distributed throughout the FoV, indicating a prolonged and widespread star formation at early epochs, as already pointed out in \citet{Sacchi2018}. 

We compared the spatial distribution of young star clusters with that of similarly aged field stars, in particular for what concerns the youngest age bin (age $<$ 10 Myr). Overall, clusters tend to align with the same large-scale features observed in the field-star population, particularly in the northern region (i.e., region 1). We identified areas, specifically to the east of region 1 (region 2) and on the southern side of the S-shaped distribution (region 3) where clusters appear in moderate abundance without a corresponding local excess of young stars. However, this discrepancy most likely result from our conservative selection of field stars, as described in Sect.~\ref{s:sp_dist_cl}. The enhancement of both clusters and field stars in region 1 appears to correlate with the elevated H~I/H$_{\alpha}$ ratio observed in that area by \citet{Hunter1996}, as well as with the excess in dust surface density observed by \citet{Calzetti2018}. This is beyond the scope of this work and will be presented in a future publication, but it is still important to highlight how combining the distributions of clusters and field stars, in conjunction with that of the neutral and ionised gas, allows to investigate key aspects such as the relative formation efficiency of star cluster vs field stars, the timescale over which clusters dissolve or become unbound, and the influence of environmental conditions on cluster formation. For instance, \citet{Bottner2003} reported that the H~I/CO ratio is higher in regions 2 and 3, which could contribute to localized differences in cluster dissolution rates, thereby affecting the observed distributions of stars and clusters across the galaxy. Finally, the overall similarity between the spatial distributions of clusters and young stars supports the idea that most, if not all, stars were initially born in clustered environments or stellar associations, although further analysis are needed to quantify this connection.  

For what concerns the AGB analysis, the CMDs reveal a noticeable gap between O-rich AGB stars and the carbon star branch, likely due to a sharp and rapid spectral transition and/or a sudden increase in carbon dust production. Additionally, a "finger" of massive AGB stars ($M_{\rm in} \geq 4 M_{\odot}$) extends vertically up to \( m_{F150W} \sim 19.7 \) mag. The height of this feature depends on factors like the IMF, star formation history (SFH) and adopted mass-loss rate, making it a crucial observational constraint on AGB evolution models.

Our results highlight how JWST’s superior resolution and sensitivity in the infrared domain open up new windows into the star formation processes of nearby dwarf galaxies. By probing the distribution of dust-enshrouded AGB stars, for instance, we can examine phases of stellar evolution that were partially or wholly missed in optical surveys. This perspective is particularly relevant for understanding how dust is generated and recycled in galaxies that experience episodic, burst-like star formation. Moreover, the identification of spatially distinct, age-sequenced star-forming regions offers a vivid demonstration of how gas dynamics, feedback, and external perturbations can sculpt the stellar content of small galaxies. 

This study serves as the first step towards a more comprehensive analysis that will include deriving the galaxy's SFH in details and comparing it to the SFH obtained in previous studies \citep[][]{McQuinn2010, Cignoni2018,Sacchi2018} in order to assess how infrared-based analyses might differ from or complement optical analyses.

%% Also note that the akcnowlodgment environment does not support long amounts of text. If you have a lot of people and institutions to acknowledge, do not use this command. Instead, create a new \section{Acknowledgments}.

\section*{Acknowledgments}
M. Correnti acknowledges financial support from the ASI-INAF agreement n.2022-14-HH-0. A.A. acknowledges support from Vetenskapsr\aa det 2021-05559. A.A and A.P. acknowledge support from the Swedish National Space Agency (SNSA) through the grant 2021- 00108. A.A. and H.F.V. acknowledges support from (SNSA) 2023-00260. This work is based on observations made with the NASA/ESA/CSA JWST. The data were obtained from the Mikulski Archive for Space Telescopes (MAST) at the Space Telescope Science Institute (STScI), which is operated by the Association of Universities for Research in Astronomy, Inc., under NASA contract NAS5-03127 for JWST. These observations are associated with
program \#1783 (PI: A. Adamo). Support for program \#1783 was provided by NASA through a grant from STScI. 
%\begin{acknowledgments}

%\end{acknowledgments}

\section*{Data Availability}
MAST data underlying this article are available at doi: http://dx.doi.org/10.17909/yrqx-fa61


\vspace{5mm}
\facilities{JWST/NIRCam}


\software{{\tt DOLPHOT} \citep[\url{http://americano.dolphinsim.com/dolphot/};][]{Dolphin2000, Dolphin2016}, {\tt Astropy} \citep{Astropy2013}, {\tt Numpy} \citep{Numpy2011}, {\tt Pandas} \citep{Pandas2010}, {\tt SciPy} \citep{Scipy2020}, {\tt Matplotlib} \citep{Hunter2007}, %{\tt APLpy} \citep{Robitaille2012, Robitaille2019}
}



%\appendix


%\bibliography{biblio}{}

\begin{thebibliography}{}

\bibitem[Ai et al.(2023)]{Ai2023}
Ai, M., Zhu, M., Xu, J., et al.\ 2023, \mnras, 524, 2911 

\bibitem[Annibali et al.(2008)]{Annibali2008} Annibali, F., Aloisi, A., Mack, J., et al. 2008, \aj, 135, 1900

\bibitem[Annibali et al.(2018)]{Annibali2018} Annibali, F., Morandi, E., Watkins, L.L., et al. 2018, \mnras, 476, 1942

\bibitem[Annibali et al.(2011)]{Annibali2011} Annibali, F., Tosi, M., Aloisi, A., et al. 2011, \aj, 142, 129

\bibitem[Annibali et al.(2012)]{Annibali2012} Annibali, F., Tosi, M., Aloisi, A., et al. 2012, \apj, 745, L1

\bibitem[Annibali et al.(2017)]{Annibali2017} Annibali, F., Tosi, M., Romano, D., et al. 2017, \apj, 843, 20

\bibitem[Astropy Collaboration et al.(2013)]{Astropy2013} Astropy Collaboration, Robitaille, T.~P., Tollerud, E.~J., et al. 2013. \aap, 558, A33, doi:10.1051/0004-6361/201322068
\bibitem[Bajaja et al.(1994)]{Bajaja1994} Bajaja, E., Huchtmeier, W. K., \& Klein, U., 1994, \aa, 285, 385

\bibitem[Bl\"ocker(1995)]{Bloecker95} Bl\"ocker, T.\ 1995, \aap, 297, 727

\bibitem[Boker et al.(2001)]{Boker2001} Boker, T., van der Marel, R. P., Mazzuca, L., at al. 2001, \aj, 121, 1473

\bibitem[Boquien et al.(2019)]{Boquien+19} Boquien, M., Burgarella, D., Roehlly, Y., et al. 2019, \aa, 622, A103

\bibitem[Bothun(1986)]{Bothun1986} Bothun, G. D. 1986, \aj, 91, 507

\bibitem[B\"ottner et al.(2003)]{Bottner2003}
B\"ottner, C., Klein, U., \& Heithausen, A. 2003, \aa, 408, 493

\bibitem[Bressan et al.(2012)]{Bressan2012} Bressan, A., Marigo, P., Girardi, L., et al. 2012, \mnras, 427, 127

\bibitem[Calzetti et al.(2015)]{Calzetti2015} Calzetti, D., Lee, J. C., Sabbi, E., et al. 2015, \aj, 149, 51

\bibitem[Calzetti et al.(2018)]{Calzetti2018}
Calzetti, D., Wilson, G.~W., Draine, B.~T., et al. 2018, \apj, 852, 106

\bibitem[Chen et al.(2019)]{Chen2019} Chen, Y., Girardi, L., Fu, X., et al.\ 2019, \aap, 632, A105. doi:10.1051/0004-6361/201936612

\bibitem[Cignoni et al.(2018)]{Cignoni2018} Cignoni, M., Sacchi, E., Aloisi, A., et al. 2018, \apj, 856, 62

\bibitem[Dell'Agli et al.(2015a)]{Dellagli15} Dell'Agli, F., Ventura, P., Schneider, R., et al.\ 2015a, \mnras, 447, 2992

\bibitem[Dell'Agli et al.(2018)]{Dellagli18} Dell'Agli, F., Di Criscienzo, M., Ventura, P., et al.\ 2018, \mnras, 479, 5035

\bibitem[Dolphin(2000)]{Dolphin2000} Dolphin, A.~E. \ 2000, \pasp, 112, 776, doi:10.1086/316630

\bibitem[Dolphin(2016)]{Dolphin2016} Dolphin, A.~E. \ 2016,  Astrophysics Source Code Library, record ascl:1608.013

\bibitem[Gelatt et al.(2001)]{Gelatt2001} Gelatt, A. E., Hunter, D. A., \& Gallagher, J. S., 2001, \pasp, 113, 142

\bibitem[Genzel et al.(1998)]{Genzel1998} Genzel, R., Lutz, D., \& Tacconi, L., 1998, Nature, 395, 859

\bibitem[Giavalisco(2002)]{Giavalisco2002} Giavalisco, M., 2002, \araa, 40, 579

\bibitem[Gregg et al.(2024)]{Gregg+24} Gregg, B., Calzetti, D., Adamo, A., et al. 2024, \apj, 971, 115

\bibitem[Kroupa(2001)]{Kroupa2001} Kroupa, P.\ 2001, 
\mnras, 322, 231. doi:10.1046/j.1365-8711.2001.04022.x

\bibitem[Hunter (2007)]{Hunter2007} Hunter, J.~D. 2007, Computing in Science Engineering, 9, 90

\bibitem[Hunter et al.(1982)]{Hunter1982} Hunter, D. A., Gallagher, J. S., \& Rautenkranz, D., 1982, \apjs, 49, 53

\bibitem[Hunter et al.(1990)]{Hunter1990} Hunter, D. A., \& Gallagher, J. S., 1990, \apj, 362, 480

\bibitem[Hunter et al.(1996)]{Hunter1996} Hunter, D. A., \& Thronson, H. A., Jr., 1996, \apj, 461, 202

\bibitem[Hunter et al.(1997)]{Hunter1997} Hunter, D. A., \& Gallagher, J. S., 1997, \apj, 475, 65

\bibitem[Hunter et al.(1998)]{Hunter1998} Hunter, D. A., Wilcots, E. M., van Woerden, H., et al., 1998, \apj, 495, L47

\bibitem[Hunter et al.(1999)]{Hunter1999} Hunter, D. A., van Woerden, H., \& Gallagher, J. S. 1999, \aj, 118, 2184

\bibitem[Karakas \& Lattanzio(2014)]{Karakas14} Karakas A.~I., Lattanzio J.~C.\ 2014, PASA, 31, e030

\bibitem[Larson \& Tinsley(1978)]{Larson1978} Larson, R. B., \& Tinsley, B. M. 1978, \apj, 219, 46

\bibitem[Le~Fevre et al.(2005)]{Lefevre2005} Le Fevre, O., et al., 2005, \aa, 439, 877

\bibitem[Lelli et al.(2014)]{Lelli2014}
Lelli, F., Verheijen, M., \& Fraternali, F. 2014, \mnras, 445, 1694

\bibitem[Marigo et al.(2013)]{Marigo2013} Marigo, P., Bressan, A., Nanni, A., et al.\ 2013, \mnras, 434, 488. doi:10.1093/mnras/stt1034

\bibitem[Marigo et al.(2017)]{Marigo2017} Marigo, P., Girardi, L., Bressan, A., et al. 2017, \aj, 835, 77

\bibitem[Marini et al. (2023)]{Marini23} Marini E., Dell'Agli F., Kamath D., Ventura P., Mattsson L., Marchetti T., Garc{\'\i}a-Hern{\'a}ndez D.~A., et al., 2023, A\&A, 670, A97

\bibitem[Martinez-Delgado et al.(2012)]{Martinez-Delgado2012} Martinez-Delgado, D., et al. 2012, \apj, 748, 24

\bibitem[McQuaid et al.(2024)]{McQuaid2024} McQuaid, T., Calzetti, D., Linden, S. T., et al. 2024, \apj, 967, 102

\bibitem[McQuinn et al.(2010)]{McQuinn2010} McQuinn, K. B. W., Skillman, E. D., Cannon, J. M., et al., 2010, \apj, 721, 297

\bibitem[Nanni et al.(2013)]{Nanni13} Nanni A., Bressan A., Marigo P., et al.\ 2013, 
 \mnras, 434, 2390

\bibitem[Paczy{\'n}ski(1970)]{Paczynski70}
Paczy{\'n}ski, B.\ 1970, \actaa, Evolution of Single Stars III. Stationary Shell Sources, 20, 287. 


\bibitem[Pastorelli et al.(2020)]{Pastorelli2020} Pastorelli, G., Marigo, P., Girardi, L., et al. 2020, \mnras, 498, 3283

\bibitem[Pedrini et al.(2024)]{Pedrini+24} Pedrini, A., Adamo, A., Calzetti, D., et al. 2024, \apj, 971, 32

\bibitem[Pettini et al.(2001)]{Pettini2001} Pettini, M., Shapley, A. E., Steidel, C. C., 2001, \apj, 554, 981

\bibitem[Pilyugin et al.(2015)]{Pilyugin2015}
Pilyugin, L.~S., Grebel, E.~K., \& Zinchenko, I.~A., 2015, \mnras, 450, 3254 

\bibitem[Reines et al.(2008)]{Reines2008}
Reines, A.~E., Johnson, K.~E., \& Goss, W. M., 2008, \aj, 135, 2222

\bibitem[Sabbi et al.(2018)]{Sabbi2018} Sabbi, E., Calzetti, D., Ubeda, L., et al., 2018, \apjs, 235, 23

\bibitem[Sacchi et al.(2018)]{Sacchi2018} Sacchi, E., Cignoni, M., Aloisi, A., et al. 2018, \apj, 857, 63

\bibitem[Sokal et al.(2015)]{Sokal+15} Sokal, K.~R., Johnson, K.~E., Indebetouw, R., et al. 2015, \aj, 149, 115

\bibitem[Stark et al. (2025)]{stark2025} Stark D.~P., Topping M.~W., Endsley R., Tang M., 2025, arXiv, arXiv:2501.17078

\bibitem[Steidel et al.(1996)]{Steidel1996} Steidel, C. C., Giavalisco, M., Pettini, M., et al.,
1996, \apj, 462, L17

\bibitem[Summers et al.(2003)]{Summers2003} Summers, L. K., Stevens, I. R., Strickland, D. K., \& Heckman, T. M., 2003, \mnras, 342, 690

\bibitem[Talent(1989)]{Talent1989} Talent, D. L., 1980, Ph.D. Thesis, Rice Univ., Houston, TX

\bibitem[Theis \& Kohle(2001)]{Theis2001} Theis, C., \& Kohle, S. 2001, \aa, 370, 365

\bibitem[Thronson et al.(1987)]{Thronson1987} Thronson, H.~A., Jr., Hunter, D. A., Telesco, C. M., et al., 1987, \apj, 317, 180
  
\bibitem[Tully et al.(2013)]{Tully2013} Tully, R.~B., Courtois, H.~M., Dolphin, A.~E., et al.\ 2013, \aj, 146, 86. doi:10.1088/0004-6256/146/4/86

\bibitem[Valdez-Gutierrez et al.(2002)]{ValdezGutierrez2002} Valdez-Gutierrez, M., Rosado, M., Puerari, I., et al., 2002, \aj, 124, 3157

\bibitem[van der Walt et al.(2011)]{Numpy2011} van der Walt, S., Colbert, S.~C., \& Varoquax, G. 2011, Computing in Science Engineering, 13, 22

\bibitem[Vassiliadis \& Wood(1993)]{vw93} Vassiliadis, E. \& Wood, P.~R.\ 1993, \apj, 413, 641, VW93

\bibitem[Ventura et al.(1998)]{Ventura98} Ventura, P., Zeppieri, A., Mazzitelli, I.,   D'Antona, F., 1998, A\&A, 334, 953

\bibitem[Ventura et al.(2012)]{Ventura12} Ventura P., Di Criscienzo M., Schneider R., 
et al.\ 2012, \mnras, 420, 1442

\bibitem[Ventura et al.(2014)]{Ventura14} Ventura P., Dell'Agli F., Schneider R., et al.\ 2014, \mnras, 439, 977

\bibitem[Ventura et al.(2018)]{Ventura18} Ventura, P., Karakas, A., Dell'Agli, F., et  al.\ 2018, \mnras, 475, 2282. 

\bibitem[Ventura et al.(2022)]{Ventura22} Ventura, P., Dell'Agli, F., Tailo, M., et al.\ 2022, Universe, 8, 45.

\bibitem[Virtanen et al.(2020)]{Scipy2020} Virtanen, P., Gommers, R., Oliphant, T.~E., et al. 2020, Nature Methods, 17, 261, doi:10.1051/0004-6361/201322068

\bibitem[Weisz et al.(2024)]{Weisz2024} Weisz, D.~R., Dolphin, A.~E., Savino, A., et al, 2024, \apjs, 271, 47

\bibitem[Wes McKinney(2010)]{Pandas2010} Wes McKinney, 2010, in Proceedings of the 9th Python in Science Conference, ed. St\'efan van der Walt \& Jarrod Millman, 56 -- 61, doi:10.1051/0004-6361/201322068

\bibitem[Whitmore et al.(2020)]{Whitmore2020} Whitmore, B. D., Chandar, R., Lee, J., et al., 2020, \apj, 889, 154

\end{thebibliography}
%\bibliographystyle{aasjournal}



\end{document}